\newif\ifNotes\Notesfalse
\newif\ifAnon\Anonfalse
\newif\ifDraft\Draftfalse
\newif\ifArxiv\Arxivfalse

\documentclass[conference]{IEEEtran}
\usepackage[hyphens]{url}
\usepackage{color}
\usepackage[table]{xcolor}
\usepackage[numbers,sort]{natbib}
\usepackage{amsmath}
\usepackage{MnSymbol}
\usepackage{wasysym}

\usepackage{booktabs} 
\usepackage{xspace}
\usepackage{hyperref}
\usepackage[flushleft]{threeparttable}
\usepackage{pifont}
\usepackage{courier}
\usepackage{graphicx}
\usepackage{array}
\usepackage{soul}
\usepackage{bm}
\usepackage{listings}
\usepackage{float}
\usepackage{subcaption}
\usepackage{oplotsymbl}
\let\labelindent\relax
\usepackage{enumitem}
\usepackage{calc}
\usepackage{placeins}
\usepackage{tikz}
\usepackage{colortbl}
\usepackage{color}
\usepackage[english]{babel}
\usepackage{blindtext}
\usepackage{marginnote}
\usepackage[nameinlink,capitalise]{cleveref}

\setitemize{noitemsep,topsep=0pt,parsep=0pt,partopsep=0pt,leftmargin=1em}

\ifDraft
  \usepackage{draftwatermark}
  \definecolor{watermarkcolor}{rgb}{0.8,0.8,1}
  \SetWatermarkText{DRAFT}
  \SetWatermarkColor{watermarkcolor}
\fi

\definecolor{linkcolor}{rgb}{0.65,0,0}
\definecolor{citecolor}{rgb}{0,0.4,0}
\definecolor{urlcolor}{rgb}{0,0,0.65}

\definecolor{cyan}{rgb}{0.0,1.0,1.0}
\colorlet{lightcyan}{cyan!60}
\colorlet{lightyellow}{yellow!60}
\colorlet{lightgreen}{green!60}

\hypersetup{hyperindex=true,pdfpagemode=UseNone,pdfstartview=FitH,colorlinks=true, linkcolor=linkcolor, 
urlcolor=urlcolor, citecolor=citecolor}

\newcommand{\swallow}[1]{}
\ifNotes
  \newcommand{\colorcomment}[2]{\leavevmode\unskip\space{\color{#1}[#2]}\xspace}
\else
  \newcommand{\colorcomment}[2]{\leavevmode\unskip\relax}
\fi

\ifArxiv

\else

\fi

\makeatletter
\newcommand\InFloat[2]{\@ifundefined{@captype}{#2}{#1}}
\makeatother

\newcommand{\marginbox}[1]{\InFloat{{\Large $\blacksquare$}}{\marginnote{\Large $\blacksquare$}}}

\newcommand{\ignore}[1]{}

\newcommand{\elmo}{\textsc{Elmo}\xspace}
\newcommand{\Elmo}{\textsc{Elmo}\xspace}
\newcommand{\Elmos}{\textsc{Elmo}*\xspace}
\newcommand{\elmos}{\textsc{Elmo}*\xspace}
\newcommand{\rosita}{\textsc{Rosita}\xspace}
\newcommand{\ttest}{$t$-test\xspace}

\newcommand{\parhead}[1]{\vspace{3pt plus 1pt minus 1pt}\par\noindent\textbf{#1}\hspace{.75em plus .5em minus .5em}}

\newcommand{\X}{$\circletfill$}
\newcommand{\A}{$\trianglepafill$}
\newcommand{\C}{$\triangleplfill$}
\newcommand{\D}{$ $}

\crefname{figure}{Figure}{Figures}
\crefname{lstlisting}{Listing}{Listings}
\Crefname{lstlisting}{Listing}{Listings}

\newenvironment{fix}{
  \renewcommand{\to}{&\ttfamily}
  \par\noindent\begin{tabular}{p{0.42\columnwidth}p{0.42\columnwidth}}\ttfamily
  }{
  \end{tabular}
}

\ifAnon
  \newcommand{\authorlist}[1]{\relax}
\else
  \newcommand{\authorlist}[1]{\author{#1}}
\fi
\newcommand{\nextauthor}{\and}
\newcommand{\myAuthor}[3]{
  \IEEEauthorblockN{#1}
  \IEEEauthorblockA{#2 \\ 
  #3\\}\\
}

\newlength{\authwidth}
\definecolor{dkgreen}{rgb}{0,0.6,0}
\definecolor{gray}{rgb}{0.5,0.5,0.5}
\definecolor{mauve}{rgb}{0.58,0,0.82}

\lstdefinelanguage[arm]{Assembler}     
{
	keywords={ldr,ldrb,str,strb,eors,rors,movs,XXX,PPP},
	comment=[l]\;
}

\newcommand{\lstdefaultset}{\lstset{
	language=[arm]assembler,
	aboveskip=5mm,
	belowskip=5mm,
	xleftmargin=0mm,
	showstringspaces=false,
	columns=flexible,
	basicstyle={\ttfamily},
	numbers=none,
	keepspaces=false,
	numberstyle=\color{gray},
	keywordstyle=\color{blue},
	commentstyle=\color{dkgreen},
	stringstyle=\color{mauve},
	breaklines=true,
	breakatwhitespace=true,
	captionpos=b,
	tabsize=3,
	frame=1	
}}

\lstdefaultset

\begin{document}

\title{\rosita: Towards Automatic Elimination of Power-Analysis Leakage in Ciphers}

\authorlist{
  \myAuthor{Madura A. Shelton}{University of Adelaide}{madura.shelton@adelaide.edu.au}
  \myAuthor{Francesco Regazzoni}{University of Amsterdam and ALaRI -- USI}{f.regazzoni@uva.nl, regazzoni@alari.ch}
  \nextauthor
  \myAuthor{Niels Samwel}{Radboud University}{nsamwel@cs.ru.nl}
  \myAuthor{Markus Wagner}{University of Adelaide}{markus.wagner@adelaide.edu.au}
  \nextauthor
  \myAuthor{Lejla Batina}{Radboud University}{lejla@cs.ru.nl}
  \myAuthor{Yuval Yarom}{University of Adelaide and Data61}{yval@cs.adelaide.edu.au}
}

\IEEEoverridecommandlockouts
\makeatletter\def\@IEEEpubidpullup{6.5\baselineskip}\makeatother
\IEEEpubid{\parbox{\columnwidth}{
    Network and Distributed Systems Security (NDSS) Symposium 2021\\
    21-24 February 2021, Virtual Conference\\
    ISBN 1-891562-66-5\\
    https://dx.doi.org/10.14722/ndss.2021.23137\\
    www.ndss-symposium.org
}
\hspace{\columnsep}\makebox[\columnwidth]{}}

\maketitle

\begin{abstract} 
  Since their introduction over two decades ago, 
  side-channel attacks have presented a serious security threat. 
  While many ciphers' implementations
  employ masking techniques to protect against such attacks, 
  they often leak secret information due to unintended interactions in the hardware. 
  We present \rosita, a code rewrite engine that uses a leakage emulator which we amend
  to correctly emulate the micro-architecture of a target system.  
  We use \rosita to automatically protect masked implementations of AES, ChaCha, and Xoodoo.
  For AES and Xoodoo, we  show the absence of observable leakage at 1\,000\,000 traces with less
  than 21\% penalty to the performance.  
  For ChaCha, which has significantly more leakage, 
  \rosita eliminates over 99\% of the leakage, at a performance cost of 64\%.
\end{abstract}

\newcommand{\nbcitep}[1]{~\citep{#1}}
\newcommand{\nbcitet}[1]{~\citet{#1}}

\section{Introduction}
The seminal work of \citet{Kocher96} demonstrated that interactions of
cryptographic implementations with their environment can result in 
\emph{side channels}, which leak information on the internal state 
of ciphers.
Since then multiple side channels have been demonstrated,
exploiting various effects, such as timing~\cite{Bernstein05,ASK05, BB03, BrumleyT11}, 
power consumption~\cite{Kocher1999},
electromagnetic (EM) emanations~\cite{QS01, GMO01, CarlierCDP04}, 
shared micro-architectural components~\cite{GeYCH18,Tromer2010},
and even acoustic and photonic emanations~\cite{KN+13, SN+13, Backes2010,GenkinST14}.
These side channels pose a severe risk to the security of systems, and
in particular to cryptographic implementations,
and effective side-channel attacks have been demonstrated against
block and stream ciphers~\cite{goubin-ches99, RO04}, public-key systems, both traditional~\cite{MDS99, ecc-dpa}
and post quantum~\cite{Park_Shim_Koo_Han_2018}, cryptographic primitives implemented in real-world devices~\cite{EisenbarthKMPSS08, BG+12}, 
and even non-cryptographic algorithms~\cite{BatinaBJP19}.

\begin{table}[t]
  \caption{Results of running \rosita to automatically fix masked implementations of AES, ChaCha, and Xoodoo.
  \label{t:summary}}
  \begin{center}
  \begin{tabular}{lrrrrr}
    \toprule
    & \multicolumn{2}{c}{\textbf{Cycles}} && \multicolumn{2}{c}{\textbf{Leakage Points}} \\
    \cmidrule{2-3}\cmidrule{5-6}
    \textbf{Function} & \textbf{Original} & \textbf{Fixed} && \textbf{Original} & \textbf{Remaining}\\
    \midrule
    AES     & 1285 & 1479 && 31 & 0 \\
    ChaCha  &  1322  & 2162      && 238 & 1 \\
    Xoodoo  &  637 &  769 && 38 & 0 \\
    \bottomrule
  \end{tabular}
  \end{center}
\end{table}

Many approaches to protect devices have been suggested, in particular 
against power and EM attacks. 
These range from special logic styles that are designed to make leakage data-independent~\cite{TV06, FG05, ChenZ06},
through noise generation to hide the signal~\cite{Mess00}, to 
algorithmic changes designed to prevent certain leakage~\cite{Mes-fse00}.
In particular, \emph{masking} is a common algorithmic countermeasure
in which all intermediate (secret-dependent) values in the ciphers are combined with random
masks, so that the leakage of one or even a few values does not provide
the attacker with enough information to recover the secrets.

The protection afforded by masking is, however, only theoretical.
In practice, masked implementations often fail to achieve the promised level of security.
One of the most common reasons for leakage from masked implementations
is unintended interactions between values in the micro-architecture.
For example, the power consumption of updating the contents
of a register may depend on a relationship between
the values prior to and after the update.

To achieve secure cryptography in the presence of side-channel attacks,
cryptographic implementations often go through multiple
cycles of leakage evaluation, 
e.g.\ as specified in International Standard ISO/IEC 17825:2016(E)~\cite{ISO17825}.
Such a process is costly because it requires a high level of expertise and
significant manual labor, especially taking into account state-of-the-art 
side-channel adversaries. 

Recently, several works have experimented with tools that 
provide a high resolution emulation of the power consumption~\cite{VeshchikovG17}.
The results of such emulations are combined with standard
statistical tests~\cite{BeckerCDGJKKLMRS13} to 
perform leakage assessment of software without executing it on the
actual hardware~\cite{Papagiannopoulos2017}.
Observing that these tools eliminate the hardware from the leakage evaluation process, 
we ask the following question:

\smallskip
\begin{center}
  \emph{Can leakage emulators be used for automatic mitigation of side channel
  leakage from software implementations?}
\end{center}
\smallskip

In this work, we answer this question in the affirmative (see \cref{t:summary} for results), albeit with some caveats.
Specifically, we develop an automatic tool, \rosita\footnote{Available at \url{https://github.com/0xADE1A1DE/Rosita}.},
that uses an emulator to detect leakage due to 
unintended interactions between values and then rewrites the
code to eliminate the leakage.
Automating leakage elimination reduces the amount of work required to ensure
adherence with the ISO~17825 standard, and to develop secure 
cryptographic implementations.

\begin{figure}[htb]
  \begin{center}
    \includegraphics[trim={12cm 0cm 7cm 1cm}, clip,width=0.7\columnwidth]{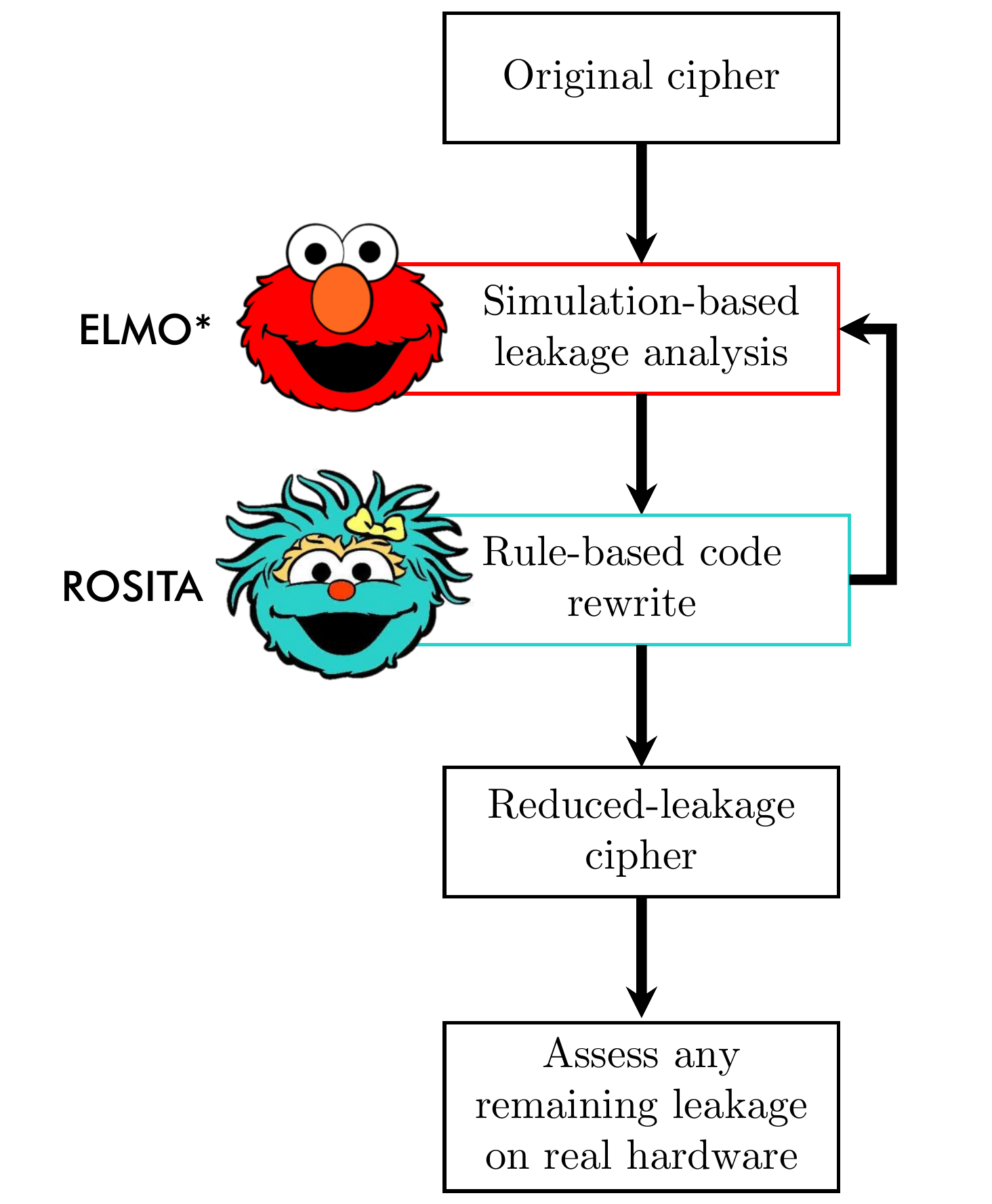}
  \end{center}
  \caption{Leakage elimination workflow with extended \Elmo (\Elmos) and \rosita.\label{f:rosita} }
\end{figure}

To emulate leakage, we develop \elmos, a leakage emulator that uses the 
execution engine of the \elmo leakage emulator~\citep{McCann2017},
but improves \elmo in three different directions.
We first note that while \elmo detects leakage between consecutive instructions,
it fails to detect leakage between instructions that are further apart.
To improve the emulation accuracy, we first design and develop
novel procedures for detecting leakage between non-consecutive instructions
and for identifying the storage components involved in this leakage.
We apply these procedures  to the Cortex M0 processor, 
identifying several storage elements.
We then modify \elmo to model these elements, achieving an accurate leakage model.
Second, we modify \elmo to not only output the instruction
that causes the leakage, but also to identify the cause of the leakage.
Last, we modify the workflow in \elmo to perform on-line statistical 
analysis of the generated traces to detect leakage.

In its core, \rosita is a rule-driven code rewrite engine.
It uses the output from \elmos to select rewrite rules and apply
them at leaky points.  
To eliminate leakage, we follow the workflow in \cref{f:rosita}.
We repeatedly execute \elmos to identify code locations that leak and then invoke \rosita to rewrite the code.
Finally, we test the code produced by \rosita on the physical device to assess the level of remaining leakage.

We experiment with masked implementations of three ciphers with very different
round functions: the AES block cipher~\cite{DaemenR02},
the ChaCha stream cipher~\cite{Bernstein08},
and the Xoodoo
permutation~\cite{DaemenHAK18}, running up to 1\,000\,000 traces on each.
We use the
table-lookup based masked implementation of AES-128 by \citet{yao2018fault}.
AES implemented with table lookups tends to be vulnerable to cache-based timing
attacks. Recent ciphers however, mostly permutations, can be implemented
efficiently with only bitwise Boolean instructions and (cyclic) shifts, e.g.,
Keccak-$p$, the permutation underlying SHA-3,~\cite{dworkin202sha,DBLP:conf/eurocrypt/BertoniDPA13},
Ascon~\cite{dobraunig2016ascon}, Gimli~\cite{bernstein2017gimli} and
Xoodoo~\cite{daemen2018xoodoo}. They all have a nonlinear layer of algebraic
degree 2 and hence allow very efficient masking. Among those, we chose Xoodoo[12]
because it is the simplest of all and it lends itself to efficient
implementations for 32-bit architectures. Finally, we consider ChaCha20 as a
symmetric-key algorithm that is very different than the other two as a representative of addition-rotation-xor (ARX) ciphers.

As \cref{t:summary} shows, \rosita successfully eliminates
leakage from the AES and Xoodoo implementations, at a moderate performance impact 
of less than 21\%.
For ChaCha, \rosita is also successful, eliminating all but one leakage points,
with a higher performance cost of 64\%.

To sum up, the contributions of this work are:
\begin{itemize}
  \item We propose a framework for generating first-order leakage-resilient implementations
    of masked ciphers by iteratively rewriting the code at leakage points.
    (\cref{s:rositaoverview})
  \item We design and implement systematic approaches  for identifying leakage through
    microarchitectural storage elements.
    We use these approaches to detect
    multiple sources of leakage in the Cortex M0 processor
    that \elmo fails to model.  We extend \elmo and augment it to
    model these sources of leakage, achieving an accurate model of leakage. 
    We further augment \elmo
    to report instructions that leak secret information and the 
    specific cause of leakage for each. (\cref{s:elmo})
  \item We develop \rosita, a code rewrite engine that uses the output of \elmos,
    our augmented \elmo, to rewrite leaking instructions and eliminate leakage.
    (\cref{s:rosita})
  \item We use \rosita to rewrite masked implementations of AES, ChaCha, and Xoodoo. We test the code
    \rosita produces and show  that \rosita eliminates almost all the leakage at
    up to 1\,000\,000 traces, with 
    an acceptable performance loss.
    (\cref{s:eval})
\end{itemize}

\section{Background}
\subsection{Side-Channel Attacks}
When software runs on hardware, it affects the environment it executes in.
This effect can be manifested as variations in power consumption,
electromagnetic emanations, temperature, and state of various CPU
components.
As these variations correlate with the operation of the algorithm, 
monitoring these variations discloses information about the internal state,
and as such provides a `communication' channel that transfers information
from the software being observed to the observer.
These unintended communication channels are often known as \emph{side channels}.

In 1996, Kocher~\cite{Kocher96} noted that the information acquired through 
a timing side channel
may reveal secret information processed by cryptographic algorithms.
Since then a significant effort has been dedicated to analyzing and eliminating side-channel
leakage, particularly in the context of cryptographic implementations~\cite{Kocher1999, QS01,GMO01,Bayrak2015,Mangard2007}.

Protection against side-channel attacks depends on both the channel
and the attacker capabilities.
The \emph{constant-time} programming style~\cite{BernsteinLS12,KS09}, 
which avoids secret-dependent branches and table lookups, has
proven effective against attacks that depend on the cryptographic
operation timing or on its effect on micro-architectural 
components~\cite{Bernstein05,BrumleyT11,GeYCH18}.
However, when the leakage correlates with the data values being processed,
constant-time programming is not sufficient to protect the implementation.

One of the main approaches to protect cryptographic implementations
against side channels that leak information on data values
is \emph{masking}~\cite{chari1999towards,Mes-fse00}.
In a nutshell, $t$-order masking represents each internal value $v$ using
$t+1$ values $v_0, v_1, \ldots, v_t$
such that the \emph{masks}
$v_1, \ldots, v_t$ are chosen uniformly at random,
and $v_0$ is set such that $v=v_0\oplus v_1\oplus\cdots\oplus v_t$.
Consequently the leakage of up to $t$ values does not disclose 
any information to the attacker,
and the implementation is secure in the $t$-probing model~\cite{IshaiSW03}.
In practice, due to the complexities involved in higher-order masking,
most masked implementation are first order, where each value
is represented by a mask and a masked value.

Although theoretically secure, naive masking often fails to provide the
required protection.
The main cause is that side-channel leakage correlates not only with the values
being processed, but also with the changes in the logical values of 
internal components, resulting in unintended interactions between values in the processor.
Past research has identified two main sources of such leakage: transitional effects and glitches.

\emph{Transitional effects} are caused when the logical value
of a register or even of a long wire, such as a bus, changes from
one to zero or vice versa.
Changing the value draws more power than maintaining the value unchanged.
Consequently, when changing the value of a register, the power consumption
corresponds with the Hamming distance between the old and new values~\cite{BalaschGGRS14}.

In contrast with transitional effects, which correspond to changes in 
logical values, \emph{glitches} are temporary changes in electrical signals
caused by signal timing differences.
Because signals take time to propagate through the circuit,
it takes time for the logical values to stabilize during a cycle.
Until the signals stabilize, they may fluctuate between signal levels,
leaking information that does not correspond just to the logical function 
computed~\cite{MangardPG05, ChenHS09}.

For masked implementations, unintended interactions, such as transitional effects
and glitches can be disastrous.
For example, suppose that a program that processes a secret value $v=v_0\oplus v_1$ contains
two consecutive instructions, where the first instruction uses $v_0$ and the second uses $v_1$.
Internally, executing the first instruction would place $v_0$ on the bus,
and executing the second instruction would change the contents of the bus
to $v_1$. 
The transitional effect of changing the contents of the bus draws power which
corresponds to the Hamming distance between $v_0$ and $v_1$, 
which is the Hamming weight of the secret value $v$.

\citet{BalaschGGRS14} demonstrate that unintended interactions 
due to transitions halve the number of intermediate values the adversary needs to acquire.
However, as mentioned above, 
algorithms that use high-order masking are significantly more
complex in terms of resources they require, than simple first-order masking.
Moreover, \citet{GaoMPO20} demonstrate that glitches may further reduce the security
level of masked implementations.

Thus, the common practice, from the practitioners' point of view, is to use first-order masking, 
and to combine it with ad-hoc countermeasures for unintended interactions.
Then, the implementation of the cryptographic software typically
undergoes leakage assessment to find code locations that
leak information. If leakage is detected, the operator applies manual modifications to 
the code to eliminate the leakage,
this process repeats until no further leakage is evident.

We now turn our attention to assessing leakage in cryptographic implementations.

\subsection{Side Channel Leakage Assessment} \label{s:scla}

Leakage assessment of a device is very important for both the semiconductor and
the security evaluation industries, and has accordingly received a lot of
attention in past years.  Depending on the attacker model, 
many attack vectors are possible and exhaustive evaluation (by trying all possible
attacks) is simply not feasible.  As an alternative, a leakage evaluation
methodology called Test Vector Leakage Assessment (TVLA) was proposed~\cite{Cooper2013, TunstallG16}.
The question that it answers relates to the presence of
any sort of leakage (from side channel measurements) of the targeted
implementation running on the device of interest. 
TVLA does have some limitations.
Specifically, a negative answer does not mean that the device is secure. However, the
confidence level can be increased by testing multiple times with different inputs.
Similarly, 
a  positive result, i.e. indication of leakage,  does not tell much on the exploitability of the leakage~\cite{SchneiderM15}.
Nevertheless, due to its
simplicity and efficacy, the TVLA method is considered useful first
diagnostics tool for side-channel leakage assessment,
and it has become the popular tool
for security analysts. 
The core idea of TVLA is to use Welch's \ttest~\cite{Welch49} to
differentiate between two sets of measurements, one with fixed inputs and the other with
random inputs. 
If the test finds sufficiently strong evidence that the measurements leak,
this implies that the device leaks
some data-dependent information through a side channel.  The main limitation is
in evaluating each point in time independently, so the leakage from combining
multiple points is not detected. To overcome this limitation,
\citet{SchneiderM15} extend the \ttest to handle multiple points. In
addition, to address leakages distributed over multiple orders they propose the
use of the ${\chi}^2$-test as a natural complement to the Welch's \ttest.

As a further guideline for analysts, the International Standard 
ISO/IEC 17825:2016(E)~\cite{ISO17825} suggests a specific procedure
for assessing the security of devices.
Specifically, the procedure requires selecting two fixed inputs
and performing TVLA measurements, comparing the traces with these
fixed inputs and those of random inputs.
The number of traces in each assessment depends on the desired
security level and ranges between 10\,000 for level 3 and 100\,000 for level 4.

\subsection{Leakage Emulators}

Because conducting real experiments for leakage detection is costly, leakage
emulation has been adapted as an alternative.  
To the best of our knowledge, PINPAS~\cite{HartogVVVW03}, which detects leakage
in Java-based smart cards, is the first such emulator.
Since then, various other methods of emulating leakage have been suggested.
Among the most accurate use SPICE~\cite{NagelP73} to simulate the internal 
circuits of a CPU down to a transistor level~\cite{AignerMMMOPST06}. 
Its drawback is that transistor-level simulators tend to be very slow.
Alternatively, researchers have looked at emulating at the source code
level~\cite{Veshchikov14} and at machine instruction
level~\cite{VeshchikovG17,Papagiannopoulos2017}. In source code level
emulation, the emulator does not have any information about a specific CPU that
will be used to run the compiled machine code of a given source code. It
emulates leakage having source code as its only input. In instruction level
emulation, the emulation is based on the machine code that will be
executed on a certain CPU or more generally a specific CPU kind. Recently,
advanced instruction level emulators have been introduced that use power and electromagnetic 
traces
from real experiments to make better estimates~\cite{SehatbakhshYZP20,McCann2017}.  Similarly,
advanced characteristics of CPUs such as instruction pipelining have found
their way into recent leakage emulators~\cite{CorreGD18}.

\textsc{Coco}~\cite{GigerlHPMB20}
suggests reformulating software testing for leakage
as a hardware verification problem.
Specifically, \textsc{Coco} 
uses a cycle-accurate simulator of masked software execution
to acquire traces of execution on a target CPU.  
It then uses \textsc{Rebbeca}~\cite{BloemGIKMW18} to verify the absence
of leakage from the underlying hardware.
An advantage of the approach is that leakage can be eliminated at
both the software and the hardware levels. 
However, it requires access to the full netlist of the target
processor and relies on manual tagging of masked values.

\subsection{Automatic Approaches to Handling Side-Channel Leakage}
Due to the numerous problems and pitfalls with countermeasures against
side-channel attacks as previously discussed, researchers developed several
automated approaches for handling side-channel leakage.  The approaches can be
grouped into three categories, simulation-based, code analysis, and hardware-assisted.

\subsubsection{Simulation-based Approaches}
\citet{Veshchikov14} presents the SILK simulator, which simulates a high level
abstraction of the source code of an algorithm that generates traces.  Another
simulator, MAPS~\cite{CorreGD18} targets the Cortex-M3 and bases its leakage
properties on the Hardware Description Language (HDL) source code.  The
simulator mainly focuses on leakage caused in the pipeline.

These two simulators only automate the generation of traces. Hence, they are
basically assist the leakage evaluation process and speed it up.

\subsubsection{Code Analysis}
\citet{BartheBDFGS15} describe how to automatically verify
higher-order masking schemes and present a new method based on program
verification techniques. The work of \citet{WangSchaumont2017}
explains how formal verification and program synthesis can be used to detect
side-channel leakage, prove the absence of such leakage and modify software to
prevent such leaks. However, both of these works remain limited in the ways
they model the hardware and actual implementations.

Closer to ours are works that, although sacrificing  generality,
address the problem of ``fixing'' the leakage from a specific device. 
\citet{Papagiannopoulos2017} perform an
in-depth investigation of device specific effects that violate the
\emph{independent leakage assumption} (ILA)~\cite{RenauldSVKF11}.
They also provide an automated tool that can
detect such violations in AVR assembly code.

Another method to eliminate timing side channels in software was proposed by
\citet{WuGuoSchaumont2018}.  Their method requires a list with secret
variables as input and produces code that is functionally equivalent to the
original code but without timing side channels.  In a recently published work
\citet{WangSW19} describe a type-based method for
detecting leaks in source code.  They implemented their mitigations in a
compiler and evaluated their method.  \citet{EldibW14} propose a
method to add countermeasures to source code that masks all intermediate
computation results such that all intermediate results are statistically
independent.

\citet{AgostaBP12} introduce a framework to
automate the application of countermeasures against Differential Power Analysis (DPA).  Their approach adds
multiple versions of the code preventing an attacker from recognizing the
exact point of leakage.


\subsubsection{Hardware-Assisted Masking}
Implementing masking within the processor promises a way of avoiding unintended
interaction between masked values.
Masking apply to the processor as a whole~\cite{GrossJMUW16,DeMulderGH19}, 
or only to a part of the instruction set~\cite{KiaeiS20,GaoGMPPR20}.

\section{\rosita Overview}\label{s:rositaoverview}
\rosita aims to automate the process of
producing leakage-resilient software.
Specifically, we focus on reducing the manual effort
required for ensuring conformance with the ISO~17825 standard.
We assume that the underlying algorithm employs a protection technique,
such as masking. 
However, unintended interactions between data, introduced 
in the execution of the software, 
can break the 
independent leakage assumption~\cite{RenauldSVKF11}
and leak secret information through
a physical side channel, such as the power channel. 

We consider two sources of interactions. In architectural interaction, the program
overwrites a register with a related value, leaking information through transitional
effects.
Microarchitectural interactions occur due to transitional effects and glitches within
the microarchitecture.  
For example, when a value of a pipeline stage register is overwritten.
We do not handle cases where the application of masking is incorrect,
either due to a programmer error or due to compiler optimizations.
Similarly, we do not protect against attacks that expose the full state of 
the cipher~\cite{KrachenfelsGMTS21}.

To fix unintended interactions, implementers typically go through a manual, iterative process
whereby the software is installed on the target device, the leakage is measured,
and fixes are applied to the machine code, until the leakage is reduced to an
acceptable level for the target use case.

This process, naturally, requires a significant level of expertise both
in setting up and conducting the experiment to assess the leakage
and in fixing the software to reduce the leakage.
Moreover, because the assessment requires a large number of encryption
rounds on relatively low-performing devices, 
and a number of repetitions in repairing the leakage and evaluating, 
the process is time consuming.

\rosita automates this process as shown in 
\cref{f:rosita}. 
To produce leakage-resilient cryptographic software,
we start with a (masked) 
implementation of the cryptographic primitive.
We use cross-compilation to produce both the assembly code 
(if the original code is in a high-level language) and the binary executable
for the target device.
The binary executable is then passed to a leakage emulator, 
in our case \elmos, a  modified version of \elmo~\cite{McCann2017}, to perform leakage assessment.
This assessment identifies the leakage and the machine instructions that cause it.
\rosita processes the output of \elmos, together with the assembly code.
It applies a set of rules that replace leaky assembly instructions 
with functionally-equivalent sequences of instructions that do not leak.
Afterwards, the produced assembly program is assembled and
fed back to \elmos and the process repeats 
until no further fixes can be applied, at which time \rosita produces
a report indicating the remaining leakage, if any.
In all of our experiments, \rosita terminates within a small number of rounds when it detects no further leakage.
The rules that \rosita applies are incremental. Hence, 
\rosita is guaranteed to converge within a finite number of rounds,
when all rules are applied in all program positions.

Note that our approach makes use of a leakage emulator. 
Prior static-analysis-based solutions,
such as~\citep{Papagiannopoulos2017, WangSchaumont2017, WuGuoSchaumont2018, Veshchikov14, CorreGD18},
rely on tags that identify the nature of values within the program.
For example, in ASCOLD~\cite{Papagiannopoulos2017} the programmer needs to assign
tags to values, e.g.\ identifying them as random or 
masked. 
The main downside of the tagging approach is that any mistake the programmer
makes in tagging values can be translated to missed leakage.
In contrast, \rosita applies TVLA, using a procedure that extends ISO~17825,
to the emulated power trace. 
As such, \rosita depends neither on the programmer's proficiency 
nor on specific properties of the masking scheme to detect leakage.
Subject to the accuracy of the emulator and the strength of the statistical tools applied, 
\rosita will detect leakage in the implementation (up to the level which the masking scheme used is meant to protect).

\section{Leakage Emulation}\label{s:elmo}
Due to \rosita's reliance on a leakage emulator,
care should be taken when selecting one.
For this work we select \elmo~\cite{McCann2017} as a basis because,
unlike instruction-level emulators, it is tailored to a specific 
processor model, while at the same time it does not require
detailed design information to build its model.
We now describe how \elmo models the device it emulates and the leakage.
We then identify limitations for using \elmo  with \rosita and describe how
we address these and develop \elmos.

\subsection{The \Elmo Leakage Model}\label{s:elmoleakagemodel}
Emulating the hardware at the transistor level would produce
the most accurate leakage estimate.
However, this is often infeasible, both due to the complexity of such
analysis and because the hardware implementation details
are not available to the security evaluators and software developers.

Instead, leakage emulators use an abstract model of the 
device and of its power consumption.
The abstract model is significantly simpler than emulating
at the transistor level. 
At the same time, using an abstract model reduces accuracy
and may result in missing some leakage.
Thus, the \emph{leakage model} presents a trade-off between
modeling cost and accuracy.

\Elmo's model of the hardware considers bit values and changes in bit values over
the Arithmetic Logic Unit (ALU) inputs and outputs and memory  instructions.
Specifically, each operand is compared to the corresponding operand of the preceding instruction. 
Power consumption is modeled as linear combinations of bit values or bit changes.

\Elmo models 21 instructions that its authors claim cover typical use in cryptography.
These 21 instructions are divided into five
groups, each modeled separately.
To generate the model, power traces are collected while
the processor executes sequences of three instructions. 
Each trace is processed to select a point-of-interest to be used
as a representative of the trace.
\Elmo then performs a linear regression on the data collected in the traces
to find the coefficients for the model.

The model itself consists of 24 main components, each modeling
a specific part of the architecture. These cover:
\begin{itemize}
  \item A linear combination of the bit flips between each operand of the
    current instruction and the corresponding operand of the previous and
    the subsequent instructions. 
  \item A linear combination of the bit values of the operands of the current 
    instruction.
  \item The instruction groups of the previous and subsequent instructions.
\end{itemize}

\Elmo provides a pre-computed model of the 
STM32F030\footnote{\url{https://www.st.com/en/evaluation-tools/32f0308discovery.html}} 
evaluation board which features
an ARM Cortex-M0 based STM32F030R8T6 
System-on-Chip (SoC).\footnote{\Elmo also provides a model for the Cortex-M4-based STM32F4 Discovery board,
which we do not use in this work.}

\subsection{Evaluation Setup}\label{s:evalsetup}
\begin{figure}[htb]
  \includegraphics[width= 0.8\columnwidth]{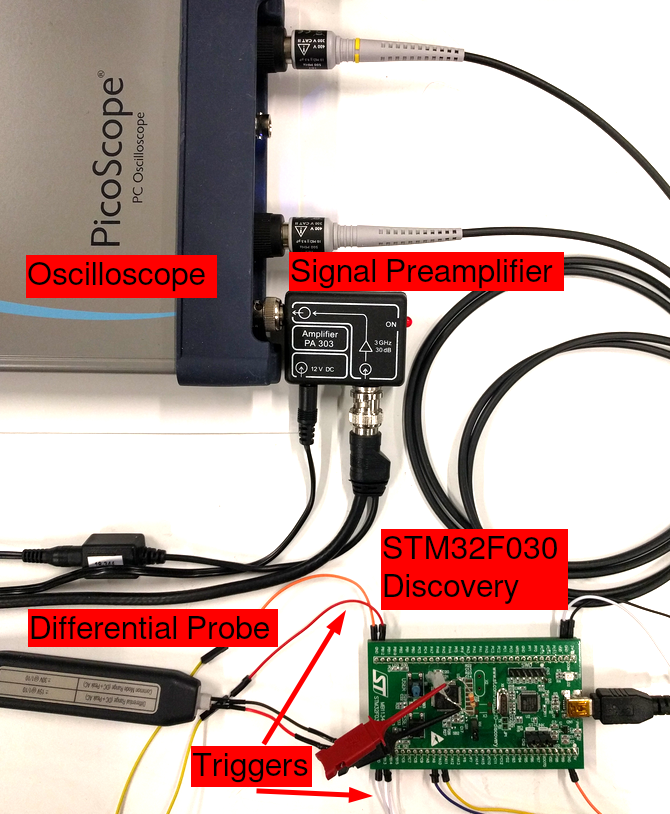}
  \caption{Evaluation Setup.\label{f:setupphoto}}
\end{figure}

To evaluate \elmo, we compare its output with leakage
assessment of the code on the real hardware.
Our evaluation setup is shown in \cref{f:setupphoto}.

We evaluate \elmo with the same 
STM32F030 Discovery evaluation board used in \citet{McCann2017}.
Following the instructions of McCann et al., 
we disconnect one of the two power inputs of the System on Chip (SoC) and attach a
330\,$\Omega$ shunt resistor to the second power input.
To avoid switching noise, we use a battery to power the evaluation board.

\begin{figure}[h]
	\centering
	\includegraphics{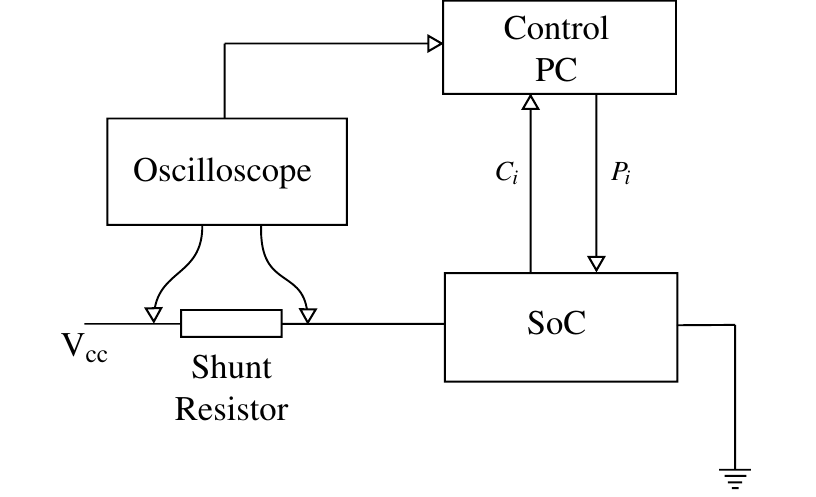}
	\caption{Evaluation Setup --- Circuit Diagram}
	\label{f:setup}
\end{figure}

We use a PicoScope 6404D with a Pico Technology TA046 differential
probe 
connected to the oscilloscope via a Langer PA 303 preamplifier,
to measure the voltage drop across the shunt 
resistor as a proxy for the power consumption of the SoC.
See circuit diagram in \cref{f:setup}.

We sample every 12.8\,ns,
which, with a clock rate of 8\,MHz, is roughly 
9.77 samples per clock cycle.
The samples are 8-bit wide 
and our PicoScope can store up to 2 giga samples
before running out of memory.

We use a control PC to orchestrate the experiments. 
The PC controls the oscilloscope and the
STM32F030 Discovery evaluation board.
It sends the software to be tested and the data to be used to the evaluation board, and
collects the trace data from the oscilloscope.
The control PC also generates all of the randomness required for the experiments.
As a source we use \texttt{/dev/urandom}, which is considered cryptographically secure.
The control PC generates the random inputs for the fixed vs.\ random tests.
It further sends a stream of random values to be used for masks by the evaluation board.

Each experiment collects multiple power traces from running the software
on the evaluation board.
The execution of the software alternates between the fixed and the random cases.
Thus, half of the collected traces are for the fixed case and the other half is
for the random tests. Fixed and random tests are run randomly interleaved to make
sure that the internal state of the device is non-deterministic at the start of
each test~\cite{SchneiderM15}. 
To identify the start and end of the segment that we monitor, 
we use the output pins of the device to trigger the trace collection and to mark
the end of the points of interest. Later, we use these trigger points to filter
out traces with clock jitter.  

To detect leakage, we employ non-specific TVLA. That is,
we check the distribution of the values at each trace point
and use the Welch \ttest to check if the samples in the fixed
and in the random traces are drawn from the same distribution.
Following the common practice in the domain, we use a \ttest value above $4.5$
or below $-4.5$ as an indication of leakage.
We validate the setup using the example from \citet{McCann2017}, getting 
results similar to theirs.
See \hyperref[app:elmovalid]{Appendix~\ref{app:elmovalid}}.

\subsection{Storage Elements and the \Elmo Model}\label{s:storage}
The \elmo model of the hardware only looks at interactions between
arguments and outputs of successive instructions. 
However, it overlooks interactions that span multiple cycles.
These interactions happen between instruction arguments
and values that are stored in storage
elements such as registers, memory, or latches.

To evaluate interactions overlooked by \elmo, we design
a systematic battery of small sequences of code that aim at highlighting interactions
via storage elements between instructions.
An example of such code is shown in \cref{l:streval}.
The code aims to check if there is an interaction between the
value stored in \cref{line:str} and the value used
as the second argument of the \texttt{eors} instruction
in \cref{line:eors}.
The purpose of the \texttt{movs} instructions between the
two tested instructions is to eliminate leakage between pipeline stages.
The sequence of nine \texttt{movs} instructions ensures that 
the \texttt{str} instruction is completed by the time the
\texttt{eors} instruction enters the pipeline.

\begin{lstlisting}[numbers=left,numberstyle=\scriptsize,keepspaces,xleftmargin=2em,caption=Evaluating interactions between the \texttt{str} and the \texttt{eors} instructions.,label=l:streval,escapechar=|]
str  r1, [r2]|\label{line:str}|
movs r7, r7
movs r7, r7
movs r7, r7
movs r7, r7
movs r7, r7
movs r7, r7
movs r7, r7
movs r7, r7
movs r7, r7
eors r3, r4|\label{line:eors}|
\end{lstlisting}

For the test, we collect 10\,000 power traces of running the code segment,
each run using different random values for the data the code processes.
(For example, in \cref{l:streval} we randomize \texttt{r1},
\texttt{r3}, \texttt{r4}, \texttt{r7}, and the contents of the memory
address pointed to by \texttt{r2}.)
For each run we also record the Hamming distance between
the two values we investigate.
(In this example, the values of \texttt{r1} and \texttt{r4}.)
We then calculate the Pearson correlation coefficient between the
Hamming distance and the values in each point of the trace.
A high correlation coefficient indicates that the Hamming distance
between the values leaks through the power trace,
implying that the first instruction keeps the value it processes in
some storage element that interacts with the data processed by the 
last instruction.

\begin{figure}[htb]
    \centering
    \includegraphics[width=\columnwidth]{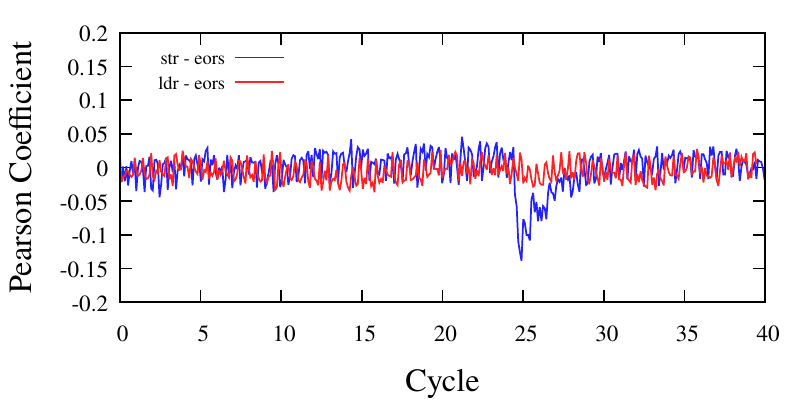}

	\caption{Pearson correlation coefficient of interference test.}
    \label{f:CPA}
\end{figure}

\cref{f:CPA} shows the Pearson correlation coefficients for two code sequences.
One from \cref{l:streval}, testing leakage from \texttt{str} to \texttt{eors},
and the other testing leakage from \texttt{ldr} to \texttt{eors}.
As we can see, the code in \cref{l:streval} show a pronounced dip in the
correlation coefficient around cycle 25, indicating interaction between the values.
Conversely, the correlation coefficient when replacing the \texttt{str}
with \texttt{ldr} remains close to zero, indicating no leakage.

%
%
%

\subsection{Dominating Instructions}
The methodology we discuss in \cref{s:storage} allows us to find instruction
pairs that interact via hidden storage within the processor.
However, each such instruction may affect multiple storage elements.
To correctly model leakage through these elements,
we need to know which instructions affect which storage elements.
Because the design details of the processor are not public, we cannot
positively identify the storage elements used by an instruction.
Instead, we search for \emph{dominating} instructions in pairs,
i.e.\ instructions that set more storage elements than others.
For that, we use code sequences similar to \cref{l:dominating},
which checks if \texttt{str} dominates \texttt{eors}.
Specifically, we pick a pair of instructions with interacting
storage.
In the test code we use two instances of the first (\cref{line:d-str,line:d-str2}), 
with the second instruction separating these two instances (\cref{line:d-eors}).
If the first instruction dominates the second, leakage will be visible at the
second instance of the first instruction (\cref{line:d-str2}).

\begin{lstlisting}[numbers=left,numberstyle=\scriptsize,keepspaces,xleftmargin=2em,caption=Checking for a dominating instruction,label=l:dominating,escapechar=|]
str  r1, [r2]|\label{line:d-str}|
movs r7, r7
movs r7, r7
movs r7, r7
eors r3, r7|\label{line:d-eors}|
movs r7, r7
movs r7, r7
movs r7, r7
str  r4, [r5]|\label{line:d-str2}|
\end{lstlisting}

Theoretically, it is possible to have a pair of instructions 
that each only affects part of the state set by the other.
However, we did not find any such pair.

\subsection{Findings}\label{s:findings}
We run a broad range of experiments, with (1) some focusing on architecturally
known storage elements, such as registers and memory, and (2) others aiming to
find micro-architectural storage elements by testing interactions between pairs
of instructions.  We find several sources of leakage that \elmo does not
identify.  We note that \citet{Gao19} also identifies many of the issues we
find; however, their identification was driven by the iterative tweaking of a
cipher, whereas our systematic approach is cipher-agnostic.
Of the leakage we find, the first is an architectural issue,
whereas the others are microarchitectural.
Except where mentioned otherwise, we believe that the cause of all microarchitectural
leakage is transition effects, because the leakage corresponds to a change in the logical
state.
However, without access to the processor implementation details, we cannot rule
out the possibility that the cause is glitches.
When the leakage does not correspond to a change in the logical state of the processor
we assume that the leakage is due to glitches.

\subsubsection{Registers}
We find that overwriting a register leaks the (weighted) Hamming distance
between the previous value and the new value.
This is a significant leakage source, because reusing a register that
contains a masked value for another value with the same mask leaks
secret information.
Unlike \citet{Papagiannopoulos2017}, we do not find leakage across different registers.

\subsubsection{Memory}
Writing data to memory interacts with data already stored in the same location.
Hence, overwriting one masked value with another may remove the mask,
leaking the values.

\subsubsection{Instruction Pairs}
We analyzed all pairs of instructions for leakage from both arguments.
The results for the second argument are summarized in 
\cref{t:leakmat1}.
We see that all instructions set some state, and that most
pairs do interact with this state. 
We now discuss some of our observations about the storage elements used.

\newsavebox{\tempboxa}
\savebox{\tempboxa}{\begin{tabular}{r@{}@{\space}l@{\space}}
		&{\footnotesize }\\
		{\footnotesize }
\end{tabular}}

\setlength{\tabcolsep}{4pt}
\begin{table}[!htb]
  \caption{State interactions between the second operands of instruction pairs. 
  Triangles point to the dominating instruction.
  Circles indicate interactions on the same storage.}
  \label{t:leakmat1}
  \resizebox{0.5\textwidth}{!}{%
    \begin{tabular}{ r| c c c c c c c c c c c c c c c c c c c c c}
      \tikz[overlay]{\draw (-4pt,5pt+\ht\tempboxa) -- (\wd\tempboxa+7pt,-\dp\tempboxa);}
      \usebox{\tempboxa} & \rotatebox{90}{eors} & \rotatebox{90}{adds} & \rotatebox{90}{ands} & \rotatebox{90}{bics} & \rotatebox{90}{cmps} & \rotatebox{90}{mov} & \rotatebox{90}{orrs} & \rotatebox{90}{subs} & \rotatebox{90}{lsls} & \rotatebox{90}{rors} & \rotatebox{90}{lsrs} & \rotatebox{90}{muls} & \rotatebox{90}{str} & \rotatebox{90}{strb} & \rotatebox{90}{strh} & \rotatebox{90}{ldr} & \rotatebox{90}{ldrb} & \rotatebox{90}{ldrh} & \rotatebox{90}{pop} & \rotatebox{90}{push}\\
      
      \hline
      eors  & \X & \X & \X & \X & \X & \X & \X & \X & \X & \X & \X & \X & \A & \A & \A & \D & \D & \D & \A & \A \\ 
      adds  & \X & \X & \X & \X & \X & \X & \X & \X & \X & \X & \X & \X & \A & \A & \A & \D & \D & \D & \A & \A \\ 
      ands  & \X & \X & \X & \X & \X & \X & \X & \X & \X & \X & \X & \X & \A & \A & \A & \D & \D & \D & \A & \A \\ 
      bics  & \X & \X & \X & \X & \X & \X & \X & \X & \X & \X & \X & \X & \A & \A & \A & \D & \D & \D & \A & \A \\ 
      cmps  & \X & \X & \X & \X & \X & \X & \X & \X & \X & \X & \X & \X & \A & \A & \A & \D & \D & \D & \A & \A \\ 
      mov   & \X & \X & \X & \X & \X & \X & \X & \X & \X & \X & \X & \X & \A & \A & \A & \D & \D & \D & \A & \A \\ 
      orrs  & \X & \X & \X & \X & \X & \X & \X & \X & \X & \X & \X & \X & \A & \A & \A & \D & \D & \D & \A & \A \\ 
      subs  & \X & \X & \X & \X & \X & \X & \X & \X & \X & \X & \X & \X & \A & \A & \A & \D & \D & \D & \A & \A \\ 
      lsls  & \X & \X & \X & \X & \X & \X & \X & \X & \X & \X & \X & \X & \A & \A & \A & \D & \D & \D & \A & \A \\ 
      rors  & \X & \X & \X & \X & \X & \X & \X & \X & \X & \X & \X & \X & \A & \A & \A & \D & \D & \D & \A & \A \\ 
      lsrs  & \X & \X & \X & \X & \X & \X & \X & \X & \X & \X & \X & \X & \A & \A & \A & \D & \D & \D & \A & \A \\ 
      muls  & \X & \X & \X & \X & \X & \X & \X & \X & \X & \X & \X & \X & \A & \A & \A & \D & \D & \D & \A & \A \\ 
      str   & \C & \C & \C & \C & \C & \C & \C & \C & \C & \C & \C & \C & \X & \X & \X & \C & \C & \C & \X & \X \\
      strb  & \C & \C & \C & \C & \C & \C & \C & \C & \C & \C & \C & \C & \X & \X & \X & \C & \C & \C & \X & \X \\
      strh  & \C & \C & \C & \C & \C & \C & \C & \C & \C & \C & \C & \C & \X & \X & \X & \C & \C & \C & \X & \X \\
      ldr   & \D & \D & \D & \D & \D & \D & \D & \D & \D & \D & \D & \D & \A & \A & \A & \X & \X & \X & \A & \A \\
      ldrb  & \D & \D & \D & \D & \D & \D & \D & \D & \D & \D & \D & \D & \A & \A & \A & \X & \X & \X & \A & \A \\
      ldrh  & \D & \D & \D & \D & \D & \D & \D & \D & \D & \D & \D & \D & \A & \A & \A & \X & \X & \X & \A & \A \\
      pop   & \C & \C & \C & \C & \C & \C & \C & \C & \C & \C & \C & \C & \X & \X & \X & \C & \C & \C & \X & \A \\
      push  & \C & \C & \C & \C & \C & \C & \C & \C & \C & \C & \C & \C & \X & \X & \X & \C & \C & \C & \C & \X \\
  \end{tabular}}
\end{table}

\subsubsection{Memory Bus}
The memory bus seems to have a storage element that stores the most
recent value stored to or loaded from the memory.
When loading from or storing to memory, the value of the storage element is
overwritten, leaking the Hamming distance between the previous and the new
value.
This leakage differs from the two described above, and happens irrespective
of the registers and the memory addresses used.
Consequently, when writing to or reading from memory, care should be taken
to only access non-secret values or values masked with different masks.
We note that the storage element could be the contents of the addressed memory
itself, where the power leakage correlates with changing the contents of the memory bus.

It is important to note that the storage element always stores a 32-bit word. 
Thus, when loading or storing a byte, the whole 4-byte aligned 
32-bit word that contains the byte
is moved to the storage element.  
This may create memory interaction between memory operations that seem completely
unrelated.
For example, consider the code in \cref{l:wordint}. In this example
we assume that memory locations 0x300 and 0x400 both contain one secret byte each, 
both masked with the same mask. 
The code in this example performs two memory operations, 
the first stores a byte into address 0x303
and the second reads a byte from location 0x402. 
We note that none of these locations contains secret data, and the data stored is
also not secret.
However, the store operation loads the 32-bit word in memory locations 0x300--0x303
into the memory bus, and the following load operation 
replaces the contents with the 32-bit word in memory location 0x400--0x403.
This causes an interaction between the data in memory locations 0x300 and 0x400,
leaking the Hamming distance between the data stored in these locations.
\begin{lstlisting}[numbers=left,numberstyle=\scriptsize,keepspaces,xleftmargin=2em,caption=Example of word interaction,label=l:wordint,escapechar=|]
movs r3, 0x303
movs r4, 0x402
movs r7, r7
movs r7, r7
movs r7, r7
strb r5, [r3]
movs r7, r7
movs r7, r7
movs r7, r7
ldrb r6, [r4]
movs r7, r7
movs r7, r7
movs r7, r7
\end{lstlisting}

A further issue in the memory bus is an interaction between the bytes of 
words loaded from or stored to memory.
Specifically, our analysis shows that when memory data is accessed, consecutive
bytes in the word interact with each other.  
Thus, if a word contains multiple bytes that are all masked with the same mask,
loading it from or storing it to memory will leak the Hamming distance
between consecutive bytes.
We note that due to the memory bus storage element described above, 
the leakage occurs even if the memory access operations access a single byte of 
a 32-bit word.

\subsubsection{Store Latch}
We find that storing a register to memory results in potential 
interactions between the value of that register and the second argument
of subsequent ALU instructions, such as \texttt{eors}.
However, if the contents of the register changes between the \texttt{str}
and the ALU instruction, the second argument of the ALU instruction
interacts with the \emph{updated} value of the register
rather than with its original value.

\begin{lstlisting}[numbers=left,numberstyle=\scriptsize,keepspaces,xleftmargin=2em,caption=Store latch example.,label=l:storelatch,escapechar=|]
str  r5, [r3]|\label{line:sl-str}|
movs r7, r7
movs r7, r7
movs r7, r7
movs r5, r2|\label{line:sl-update}|
movs r7, r7
movs r7, r7
movs r7, r7
eors r1, r4|\label{line:sl-eors}|
\end{lstlisting}

For example, in the example in \cref{l:storelatch}, the code stores
the value of \texttt{r5} to memory (\cref{line:sl-str}).
It then updates the value of \texttt{r5}, moving the
contents of \texttt{r2} to it (\cref{line:sl-update}).
Finally, it calculates the exclusive-or of \texttt{r1} and \texttt{r4}.
Our experiments show leakage in \cref{line:sl-eors},
which correlates with the Hamming distance between the original values
of \texttt{r2} and \texttt{r4}.
Interestingly, we note that the update of the interacting register takes
one cycle to become effective.  
That is, removing Lines~6--8 in \cref{l:storelatch} removes the
interaction between the original values
of \texttt{r2} and \texttt{r4}, but leaves an interaction between
the original values of \texttt{r5} and \texttt{r4}.

We believe that the processor maintains a reference
to the most recently stored register.
This reference is used as an input to a multiplexor that selects the
contents of the referenced register.
Implementing the \texttt{str} instruction
requires two cycles~\cite[Figure~4.6]{Furber00}.
In the first, the processor calculates the store address
and in the second it performs the store.
We believe that, to avoid locking the register file,
in the first cycle the processor copies the contents of the register 
to an intermediate latch, from which it is retrieved in the second
cycle.
We believe that a glitch on the bus causes interference
between the contents of the latch and the second argument
of subsequent instructions, explaining the leakage we observe.

\subsection{Extending the \Elmo Model}\label{sec:elmos}
Recall (\cref{s:elmoleakagemodel}) that \elmo builds its model using a linear
regression from traces collected from sequences of three instructions.
To account for the effects of storage elements we identified,
we update the model to include a few more components. We call out extension \Elmos.

Whereas the \elmo model treats each operand separately,
we also look at combinations of bits across the two operands of
the instruction.
Because the first operand is typically the destination register,
correlating the two operands captures the effect of calculating
the result of the operation and overwriting the destination register.

To capture interactions via memory and internal storage elements,
we track the contents of these elements, and add model components that
correlate with them.

\begin{figure}[htb]
	\centering

	\includegraphics[width=\columnwidth]{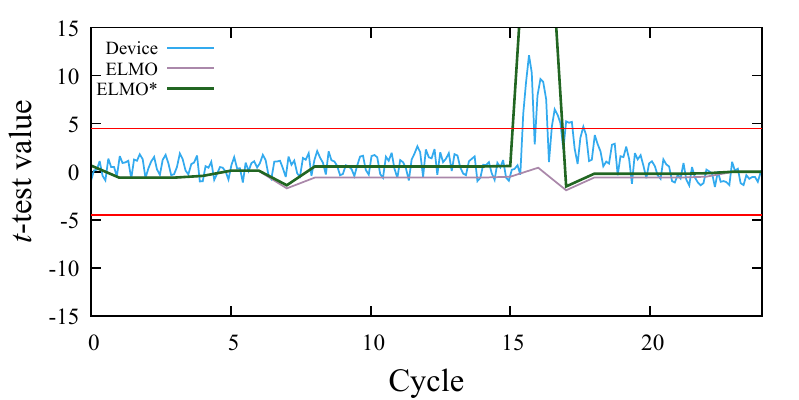}
		
	\caption{Leakage from \texttt{str} to \texttt{eors}.\label{f:str-eors}}
	
\end{figure}

In total, our model consists of  25 components. 
We validate our model by repeating the test cases used for identifying the storage
elements.
For example, \cref{f:str-eors} shows the real and the emulated leakage from
running the code in \cref{l:streval} on the real hardware,
\elmo, and in \elmos.
We see that our model identifies the leakage
that the original \elmo model misses.
We note that the leakage in \elmos has a higher $t$-test value than the actual leakage.  We speculate that the reason is that the model does not suffer from physical noise.
As a result, \elmos requires less traces than a hardware measurement for leakage detection.

Unlike \elmo, \elmos  not only emulates
the leaked signal, but also finds which of the components of the
model is causing the leak.
We, therefore calculate the \ttest value not only for the combined 
emulated signal, but also to separate components of the signal.\footnote{Implementation note: to reduce memory usage, we calculate the \ttest values
incrementally, using Welford's
algorithm~\cite{Welford62}.}
Thus, for example, we keep track of the \ttest value of the part
of the signal contributed by each of the instruction operands,
by interactions between the instruction result and its operands,
and by interactions between the instruction results or operands with
values previously stored to or loaded from memory.
Using this information, when \elmos reports that an instruction leaks,
we can inspect the components and identify the leakage cause.

\section{Code Rewrite in \rosita}\label{s:rosita}
As \cref{s:rositaoverview} describes, the core of \rosita is a rewrite engine
that uses the output of the \elmos emulator to drive code fixes for leakage.
We assume that the original code is masked, i.e.\ it does not leak at the
algorithm level.  
However, the translation of the algorithm into machine code and the execution
of this machine code can result in unintended and unexpected leakage.
In this section, we review the causes of leakage we identify, and
describe the fixes the \rosita applies for each.
We begin with a high-level description of \rosita and proceed
with the details of the rewrite rules it applies.

\subsection{\rosita Design}\label{s:rdesign}
\rosita is a rewrite engine that takes the code and the output of \elmos,
and rewrites the code to avoid leakage.
To decide which rewrite rule to apply, \rosita relies on
\elmos to identify the leaking component. (See \cref{sec:elmos}.)

The main strategy \rosita uses to fix the leakage is to wipe stored state
with a random mask. 
For that, \rosita dedicates a \emph{mask register}
(\rosita uses register \texttt{r7}),
which is initialized with a random 32-bit mask.
When compiling the software, we use the flag \texttt{-ffixed-r7} 
to direct the compiler not to use the
mask register, ensuring that its contents are not modified except by
\rosita.
Similarly, we require assembly implementation to not use \texttt{r7}.

It is important to note that all of the rewrite rules do not eliminate
values used by the program.
Thus, if the implementation is not fully masked or uses incorrect randomness,
\rosita will be unable to remove the leakage.
Conversely, a success in eliminating a leak is a demonstration that
the leak originated from unintended interaction.

\subsection{Operand Interaction}
One of the common forms of unintended interaction is between
the operands of successive instructions.
Technically, as \citet{McCann2017} note, loading an operand to the
bus 
leaks the Hamming
distance between
the value previously held in the bus and the new value.
If both values use the same mask, the Hamming distance
between the masked values is the same as that between the original values.

\rosita identifies such leakage by checking the various \ttest values calculated for
the operands and their relationship with those of prior instructions.
In the case that the leakage is caused by such interaction, 
\rosita inserts an access to the mask register, using
\texttt{movs r7, r7} 
The instruction moves the contents of the mask register into the mask register,
and is therefore functionally a no-op. 
However, because the value of the mask register goes through the bus, the previous
contents of the bus is wiped, removing the interaction between the two masked values.

\subsection{Register Reuse}
Due to the limited number of registers, compilers and programmers
often reuse those,
e.g.\ when the old contents are either
consumed or stored in memory.
Reusing a register rarely removes the old contents from it.
Consequently, when new data is loaded into a register, 
it interacts with algorithmically unrelated data that remains
from prior uses of the register.

\citet{Papagiannopoulos2017} note that if the old contents and the new
contents are both masked using the same mask value, the difference
between the masked contents, i.e.\ their exclusive or, 
is the same as the difference between the unmasked contents.
Consequently, when a register is used consecutively for two values with the same mask,
it leaks the difference between the values.

To identify this form of leakage, \rosita checks the \ttest value of overwriting
register value. 
Once identified, \rosita wipes the old contents of the register by copying the 
contents of the mask register to the destination register of the leaking
instruction, as \citet{Papagiannopoulos2017} suggest.
For example, suppose that  the instruction \texttt{movs r3, r4} leaks because
both \texttt{r3} and \texttt{r4} contain values masked with the same mask.
To eliminate the leak, \rosita inserts \texttt{movs r3, r7} before the leaking instruction.

\subsection{Rotation Operations}
Rotation operations show interaction between the value pre and post
rotation.
When a single masked value is rotated, this interaction is unlikely to leak
secret data because the mask hides the contents.
However, when rotating a word comprised of multiple masked values
that all use the same mask, the result of the rotation may align the
masked values, effectively nullifying the mask, leaking the difference of the
unmasked values.

We propose two approaches to remove this leakage.
As an example, suppose that we would like to rotate the register
\texttt{r2}, whose value is a concatenation of four masked bytes:
$(b_1\oplus m) || (b_2\oplus m) || (b_3\oplus m)||(b_4\oplus m)$.
Rotation of \texttt{r2} by a multiple of 8 bits would result in 
leakage of information on the value of the $b_i$'s. 
For example, assuming \texttt{r3} contains the value 8,
the instruction \texttt{ror r2, r3} would set the value of \texttt{r2}
to $(b_2\oplus m) || (b_3\oplus m)||(b_4\oplus m)||(b_1\oplus m)$,
and the interaction between the original and the rotated values of \texttt{r2}
would leak the Hamming weight of
$(b_1\oplus b_2) || (b_2\oplus b_3) || (b_3\oplus b_4)||(b_4\oplus b_1)$.

\parhead{Word Mask.}
A straightforward approach for preventing such leakage
is to mask the word with our mask register (\texttt{r7}),
rotate both the word and the mask register and then use the rotated mask to unmask the word.
Thus, instead of rotating \texttt{r2}, we rotate $\texttt{r2}\oplus \texttt{r7}$.
As an example, \cref{f:rormask} shows how \rosita fixes a \texttt{rors r2, r3}
instruction that \elmos indicates is leaking. 

\begin{figure}[htb]
  \begin{centering}
    \begin{fix}
      rors r2, r3
      \to
      eors r2, r7\linebreak
      rors r2, r3\linebreak
      rors r7, r3\linebreak
      eors r2, r7 
    \end{fix}
  \end{centering}
  \caption{Masking rotation operations. The leaking \texttt{ror} operation on the left
  is replaced with a masking sequence on the right.\label{f:rormask}}
\end{figure}

We note that this sequence modifies the contents of our mask register.
However, this has no effect on the functionality because the mask register is
assumed to be random and there is no long-term dependency on its exact contents.

\parhead{Partial Rotations.}
An alternative approach is to combine multiple shifts to avoid rotations
of multiples of the data size. 
For example, a rotation by 8 bits can be replaced with a rotation by 3 bits followed
by a rotation by 5 bits.

\rosita employs the word mask approach both because it is more general, 
i.e.\ does not depend on the size of the rotation,
and because it already has the mask register, which it uses for the other fixes.

\subsection{Memory Operations}\label{s:rewritememory}
As discussed in \cref{s:findings}, 
there are several effects that can cause interactions between values used
in memory operations.
These include a storage element in the memory bus that remembers
recently accessed memory value and consequently leaks the Hamming distance
between the remembered value and the current one on memory access operations,
interaction between loaded and stored values and the previous contents they overwrite,
and an interaction between bytes in stored words.

When \elmos indicates that a load instruction leaks due to interaction with the memory bus
storage element, \rosita wipes the contents of the bus
by pushing the mask register to the stack and popping from the
stack to the destination register of the load instruction.
\cref{f:loadfix} shows an example of an \texttt{ldr} instruction (left) 
that leaks through interaction of the loaded value with a previously loaded value.
To fix this, \rosita inserts a \texttt{push} and a \texttt{pop} instructions before 
the load, yielding the code fragment in the right.
Popping the mask to the destination of the load instruction also protects
against leakage through interaction with the previous value of the destination
register.

\begin{figure}[htb]
  \begin{centering}
\begin{fix}
    ldr r2, [r3]
  \to
    push r7\linebreak
    pop r2\linebreak
    ldr r2, [r3]
\end{fix}
  \end{centering}
  \caption{A leaking load instruction (left) and the fixed sequence (right).\label{f:loadfix}}
\end{figure}

Due to the more intricate potential interactions, 
the picture with store instructions is a bit more complex. 
To overcome interactions with the previous value used on the memory
bus and to address possible interactions with the previous contents
of memory, \rosita first stores the mask register into the destination 
location and then performs the required store
(See \cref{f:storefix}).

\begin{figure}[htb]
  \begin{centering}
\begin{fix}
    str r2, [r3]
  \to
    str r7, [r3]\linebreak
    str r2, [r3]
\end{fix}
  \end{centering}
  \caption{A leaking store instruction (left) and the fixed sequence (right).\label{f:storefix}}
\end{figure}

When byte interaction within the stored data leaks, \rosita stores
one byte at a time.  
In such a case, care should be taken to ensure that these bytes and the
operations required for their storage do not create unintended interactions,
leading to a relatively long code segment in 
\cref{f:storebytefix} in \cref{app:byteinteract}.
While this rewrite rule eliminates the leakage, the performance cost of using it
is significant. 
As such, it may be better to avoid stores of words that contain multiple values masked with
the same mask.
Changing the logic of the cipher is outside the scope of \rosita.

\begin{figure*}[htb]
  \begin{subfigure}{0.32\textwidth}
    \centering
    \includegraphics[width=0.9\textwidth]{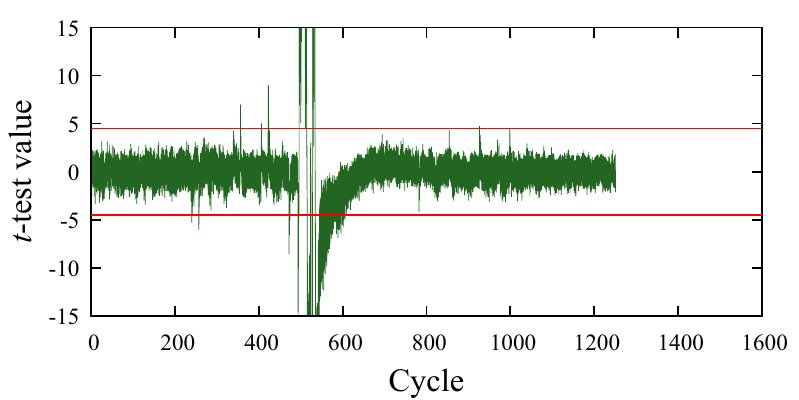}
    \caption{AES original implementation.}
    \label{f:aes_check_200k}
  \end{subfigure}
  \begin{subfigure}{0.32\textwidth}
    \centering
    \includegraphics[width=0.9\textwidth]{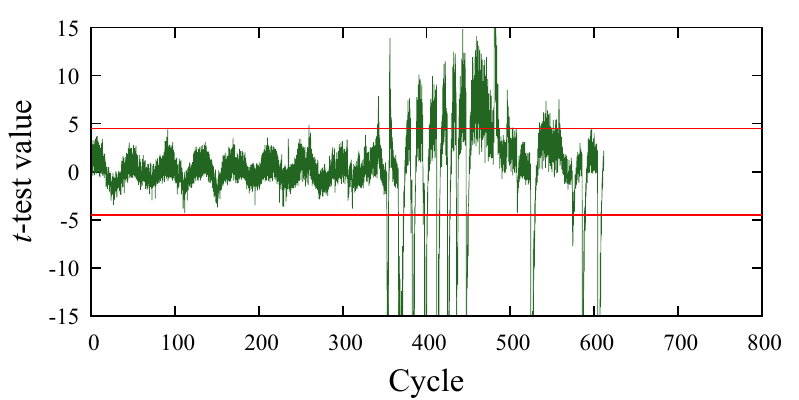}
    \caption{Xoodoo original implementation.}
    \label{f:xoodoo_check_200k}
  \end{subfigure}
  \begin{subfigure}{0.32\textwidth}
    \centering
    \includegraphics[width=0.9\textwidth]{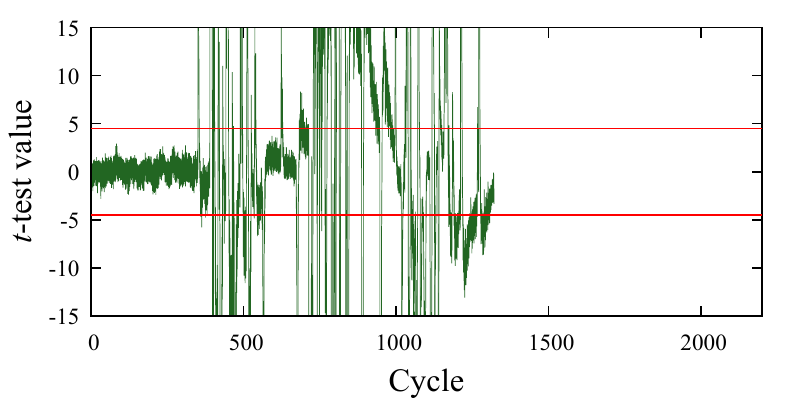}
    \caption{ChaCha original implementation.}
    \label{f:chacha_check_200k}
  \end{subfigure}

  \begin{subfigure}{0.32\textwidth}
    \centering
    \includegraphics[width=0.9\textwidth]{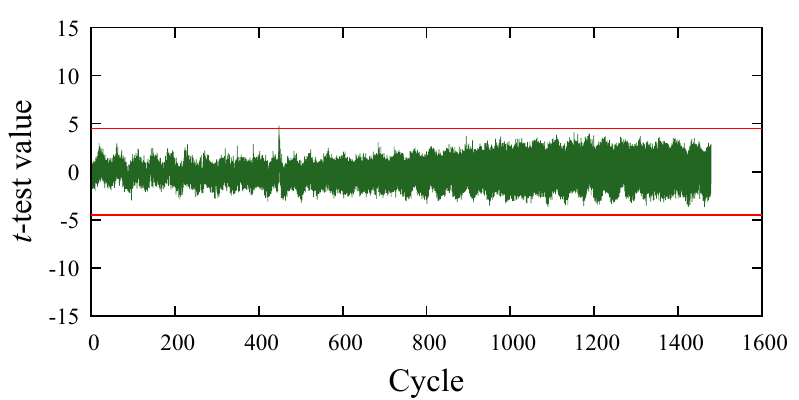}
    \caption{AES fixed with \rosita.}
    \label{f:aes_fixed_200k}
  \end{subfigure}
  \begin{subfigure}{0.32\textwidth}
    \centering
    \includegraphics[width=0.9\textwidth]{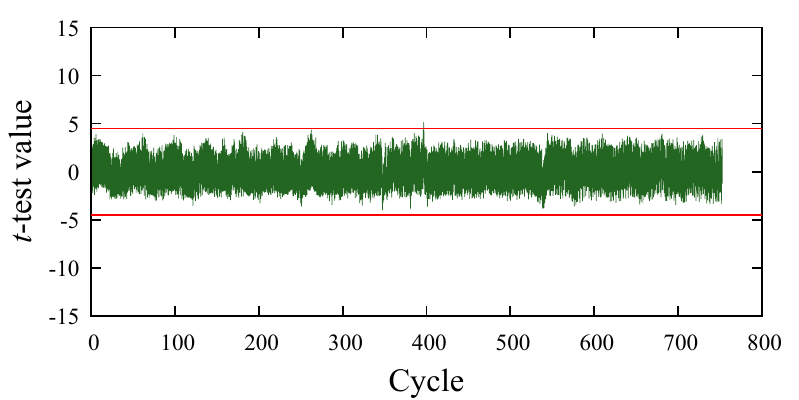}
    \caption{Xoodoo fixed with \rosita.}
    \label{f:xoodoo_fixed_200k}
  \end{subfigure}
  \begin{subfigure}{0.32\textwidth}
    \centering
    \includegraphics[width=0.9\textwidth]{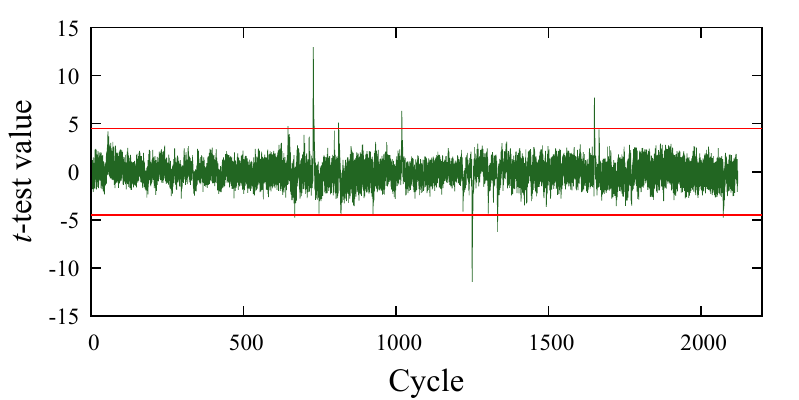}
    \caption{ChaCha fixed with \rosita.}
    \label{f:chacha_fixed_200k}
  \end{subfigure}

  \caption{Fixed vs.\ random tests for the three cipher (one fixed input, 1M traces).}
  \label{f:fixedvsrandom}
\end{figure*}

\section{Evaluation} \label{s:eval}
We evaluate \rosita with masked implementations of three cryptographic
primitives.
AES~\cite{DaemenR02} is one of the most commonly used ciphers, having been an international
standard since 2001.
We use the byte-masked 
implementation of AES-128 by Yao et al.~\cite{yao2018fault}.\footnote{\url{https://github.com/Secure-Embedded-Systems/Masked-AES-Implementation/tree/master/Byte-Masked-AES}}
To perform the \textsc{ShiftRows} operation of AES, 
which permutes bytes in the data being encrypted,
the implementation uses byte loads and stores.
Following the suggestion of \citet{Gao19}, we use different masks for each row
to avoid leakage through interactions between bytes in memory words.

Efficient software AES implementations on CPUs (without dedicated AES
instruction) use table lookups, which makes them vulnerable to cache-based
attacks. As an example of a modern primitive, the second cipher we use is the
cryptographic permutation Xoodoo.  Xoodoo was proposed recently by Daemen et
al. in~\cite{DaemenHAK18} for use in authenticated encryption
modes~\cite{daemen2018xoodoo}. The optimized and non-masked implementation of
Xoodoo we took from~\cite{bertoniextended} and we implemented the 2-share boolean
masking scheme of the non-linear layer $\chi$ as in~\citet{KeccakImplementation} ourselves.
Implementing the 2-share boolean masking of the linear layer was trivial.
In contrast to what~\citet{KeccakImplementation} mention, we initialize the state with
fresh randomness for each trace to keep it consistent with the implementation of AES,
even though this is not required.

The third cipher we consider is ChaCha as a prominent example of an ARX cipher.
ChaCha is very efficient in software and widely used in TLS implementations.
The challenge is to mask it at low cost and the best result for ARM Cortex-M3 and Cortex-M4 processors was recently
published by Jungk et al.~\cite{jungk2018efficient}. We use their implementation in our experiments.

\subsection{Fixing Leakage}

We first show \rosita's success in fixing the leakage it detects.
\cref{f:aes_check_200k} shows the results of a non-specific fixed vs.\ random
experiments with 1\,000\,000  traces of executing the first round of the AES implementation.
The figure shows leakage ($t$-value above the threshold of 4.5) around cycles
500--550, which correspond to the AES \textsc{ShiftRows} operation. 
As \cref{f:aes_fixed_200k} shows, \rosita detects the leaks and fixes them.
This fix does not, however, come for free.  The first round now takes 1\,479 cycles,
compared with 1\,285 for the original implementation---a slowdown of 15\%.
\cref{f:xoodoo_check_200k,f:xoodoo_fixed_200k} show similar results for ChaCha and Xoodoo with 61\% (1322 vs.\ 2122 cycles) and 18\% (637 vs.\ 753 cycles) slowdowns respectively. 

\begin{figure}[htb]
	\centering
	\includegraphics[width=\columnwidth]{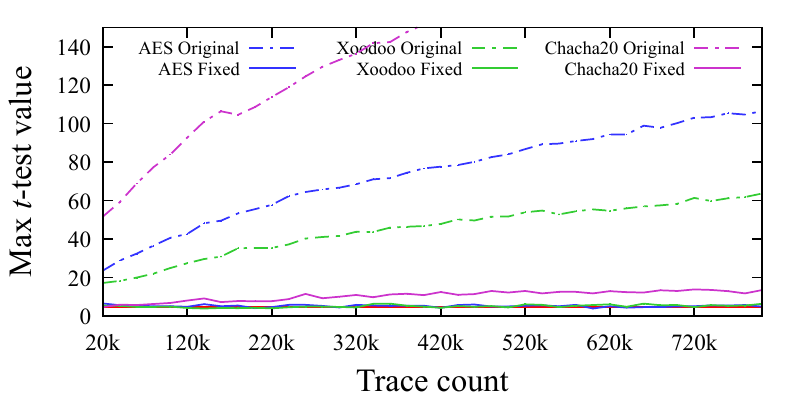}

	\caption{\ttest value trend.}
	\label{f:ttest_trend}
\end{figure}

To determine the trend of leakage, we perform the fixed vs.\ random test on the
hardware with a varying number of traces.
\cref{f:ttest_trend} shows the results for both the original and the fixed
implementations. 
The horizontal axis shows the number of traces used for the fixed vs.\ random test,
and the vertical is the maximum absolute value of the \ttest for each of the implementations.
As we can see, the original implementations show increasing leakage,
significant leakage is visible even with as little as 1\,000 traces,
and the confidence increases as traces are added.
Our fixed implementations show significantly less leakage up to 1\,000\,000 traces.
To remove the remaining leakage, we need to use more than one input.
We discuss this issue next.

\subsection{Multiple Fixed Inputs}

\begin{figure*}
  \begin{subfigure}{0.32\textwidth}
    \centering
    \includegraphics[width=0.9\textwidth]{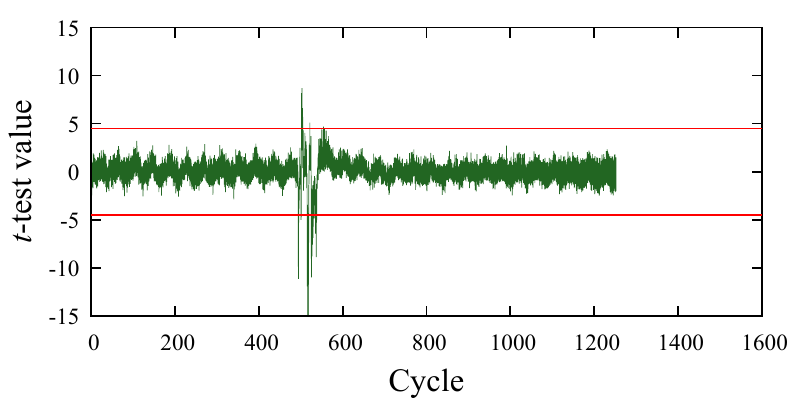}
    \caption{One fixed input -- before \rosita.\label{f:aes-1b}}
  \end{subfigure}
  \begin{subfigure}{0.32\textwidth}
    \centering
    \includegraphics[width=0.9\textwidth]{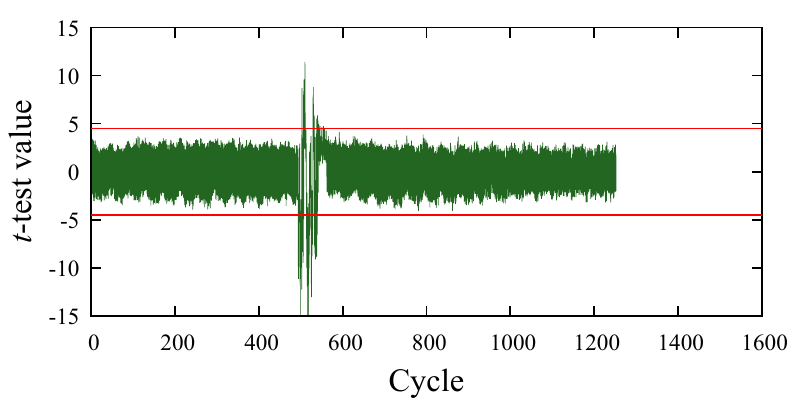}
    \caption{Five fixed inputs -- before \rosita.\label{f:aes-5b}}
  \end{subfigure}
  \begin{subfigure}{0.32\textwidth}
    \centering
    \includegraphics[width=0.9\textwidth]{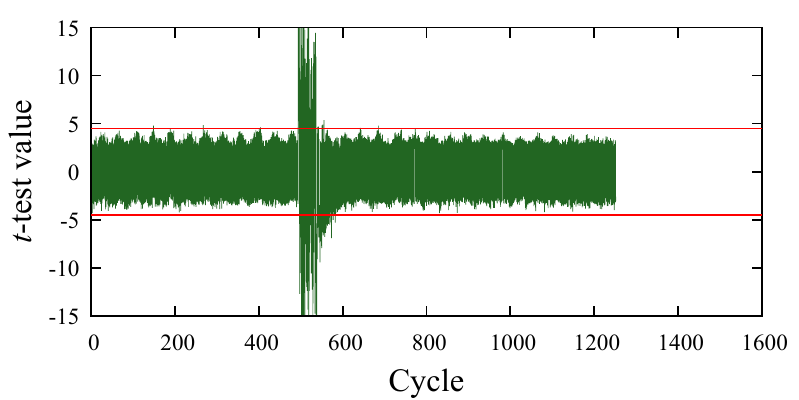}
    \caption{100 fixed inputs -- before \rosita.\label{f:aes-100b}}
  \end{subfigure}\\
  \begin{subfigure}{0.32\textwidth}
    \centering
    \includegraphics[width=0.9\textwidth]{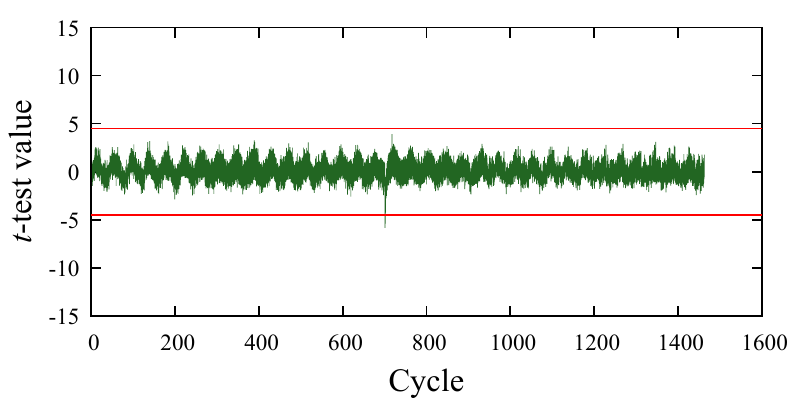}
    \caption{One fixed input -- after \rosita.\label{f:aes-1a}}
  \end{subfigure}
  \begin{subfigure}{0.32\textwidth}
    \centering
    \includegraphics[width=0.9\textwidth]{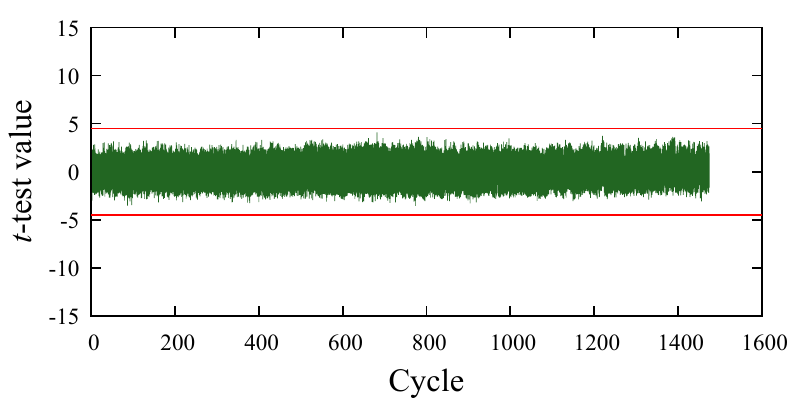}
    \caption{Five fixed inputs -- after \rosita.\label{f:aes-5a}}
  \end{subfigure}
  \begin{subfigure}{0.32\textwidth}
    \centering
    \includegraphics[width=0.9\textwidth]{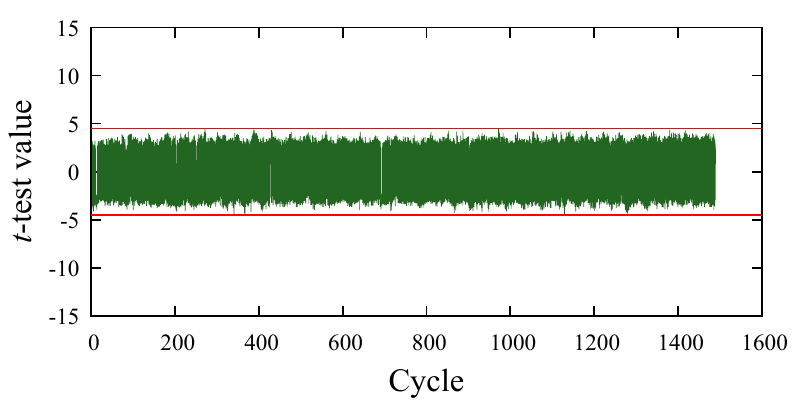}
    \caption{100 fixed inputs -- after \rosita.\label{f:aes-100a}}
  \end{subfigure}
  \caption{$t$-test of masked AES implementation before and after \rosita, varying the number of fixed vs.\ random pairs.\label{f:AES-fix}}
  \vspace{-1em}
\end{figure*}

\begin{figure*}[htb!]
  \begin{subfigure}{0.32\textwidth}
    \centering
    \includegraphics[width=0.9\textwidth]{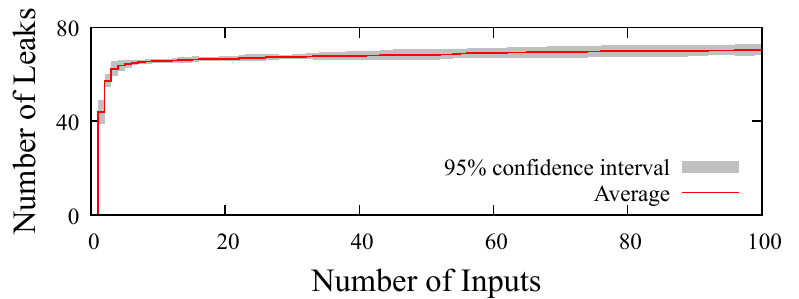}
    \caption{AES}
    \label{f:aes_100_avg}
  \end{subfigure}
  \begin{subfigure}{0.32\textwidth}
    \centering
    \includegraphics[width=0.9\columnwidth]{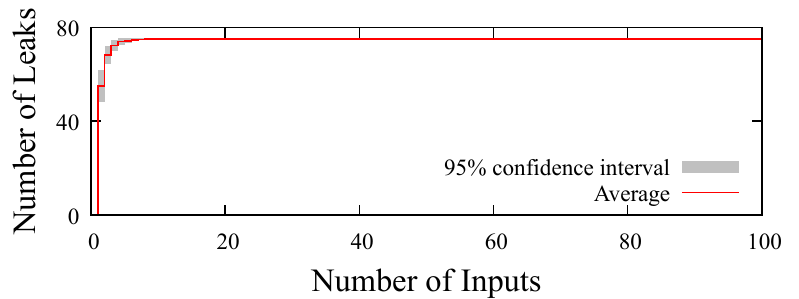}
    \caption{Xoodoo}
    \label{f:xoodoo_100_avg}
  \end{subfigure}
  \begin{subfigure}{0.32\textwidth}
    \centering
    \includegraphics[width=0.9\columnwidth]{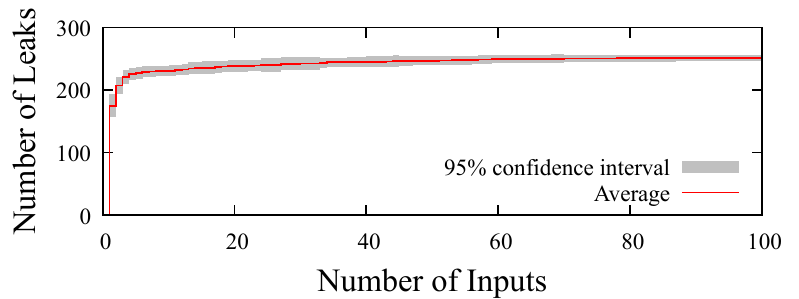}
  	\caption{ChaCha}
  	\label{f:chacha20_100_avg}
	\end{subfigure}
  \caption{Average number of leaks per fixed inputs.}
  \label{f:leaktrend}
  \vspace{-1em}
\end{figure*}

One of the known limitations of TVLA is that it may miss some leakage if used
with only one input~\cite{standaert2018not}.
Thus, common operating procedures require running multiple fixed vs.\ random tests,
each with a different fixed input.
For example, the ISO~17825 standard requires two fixed inputs.
To test the impact of multiple fixed inputs,
we perform multiple fixed vs.\ random tests, each with a different
randomly chosen fixed input.
To combine the results of multiple experiments, 
we use the largest absolute $t$-value calculated for a sample point as a representative
for leakage at that point.
Thus, if any of the fixed vs.\ random tests indicates leakage at a point, 
the combined result will also indicate leakage there.
The top row of \cref{f:AES-fix} shows the combined results
from 1, 5, and 100 fixed inputs for AES.
As we can observe, when we increase the number of fixed points, the number of 
locations that show leakage increases.
\cref{f:Xoodoo-fix,f:chacha20-fix} in \cref{app:figures} show the
results for Xoodoo and ChaCha.

Running \rosita with these inputs allows us to fix most of these leakages.
Repeating the experiment with the code that \rosita produces and the
same fixed inputs we get the \ttest values in the
bottom rows of the figures.
We can observe that \rosita fixes the leakage, at a cost to the performance. 
Also, when there is higher leakage points, it is required to have more fixed
inputs to cover all of them. For ChaCha, 100 fixed inputs were
not enough to cover all leakage points as the final result shows 1 leaky point.
\cref{t:overhead} compares the performance of the code before and after
running \rosita, showing a maximum overhead of 64\% for ChaCha.

\begin{table}[htb]
  \caption{Encryption length (cycles) after fixing with \rosita with varying number of fixed inputs.\label{t:overhead}}
  \centering
  \begin{tabular}{lrrrr}
    \toprule
    \textbf{cipher}& \textbf{Original} & \textbf{1 Fixed} & \textbf{5 Fixed} & \textbf{100 Fixed}\\
    \midrule
    AES & 1\,285 & 1\,464 & 1\,475 & 1\,479\\
    ChaCha & 1\,322 & 2\,066 & 2\,114 & 2\,162\\
    Xoodoo & 637 & 735 & 765 & 769\\
    \bottomrule
  \end{tabular}
\end{table}

The number of leakage points identified depends on the fixed inputs chosen for the 
fixed vs.\ random test.
To better understand the relationship, we want to find out how many 
leakage points we expect to find for a given number of fixed vs.\ random inputs.
\cref{f:leaktrend} shows this for AES, ChaCha and Xoodoo.
For a given number of inputs, the plots display the average number of
leakage points that \elmos discovers over 10 selections of inputs.
The figure also displays the 95\% confidence interval.
We see that 10 fixed inputs are enough to find 93\% of the leakage points in AES, 92\% in ChaCha and 99\% in Xoodoo.
When we compare the identified leakage points against the ground truth, we find
that many of the discovered leakage points are false positives, 
explaining the discrepancy between the figures and \cref{t:summary}.
To verify that we discover all of the real leakage points,
we use \rosita with 100 fixed inputs to fix AES and Xoodoo.
We then use the produced code on the hardware with
a new set of 100 fixed inputs.
We found no evidence for leakage with either AES or Xoodoo but for ChaCha we
found one leaky point that had a $t$-test value of $5.2$.

We note that \rosita's success in eliminating leakage demonstrates that
the original programs are, indeed, first-order secure, at least semantically.
As we mention in \cref{s:rdesign}, the rewrite rules can only fix leakage that stems from 
unintended interactions.

\subsection{Performance}
The performance bottleneck for \rosita is running \elmos to generate 
the simulated traces. 
We can collect 10\,000 traces of AES in 26 seconds, ChaCha in 45 and Xoodoo in 21.
In comparison, the code rewrite phase of \rosita takes around 0.1 seconds.

Collecting the same number of 1-round traces from the hardware takes
117 seconds for AES, 220 for ChaCha and 147 for Xoodoo.
The collection of traces for Xoodoo from the hardware is slower than for AES
because we provide fresh shares for every trace as mentioned above. 
Hence, the communication dominates the execution time.
Thus, \elmos is 4.5--7 times faster than the real hardware.

We note that the task of collecting traces is ridiculously parallelizable.
Hence, on a typical desktop, we can collect traces eight times faster,
and with an investment of \$1\,000 we can double the rate again.
In contrast, to parallelize trace collection from the hardware,
we would need to replicate the setup, at a cost of over \$10\,000 per node.
Thus, the effective speed of \rosita is about two order of magnitude faster 
than the hardware.

%


\section{Limitations and Future Work}
Possibly the main limitation of \rosita is that,
while we have found no evidence for leakage, 
there is no guarantee that it fixes all leakage.
There two main reasons for that:
\begin{itemize}
  \item \textbf{Methodology:} While popular and standardized, non-specific fixed vs.\ random
    tests are not a panacea for leakage. 
    Leakage they detect is not necessarily exploitable
    and the absence of detection does not necessarily mean that no leakage exists.
    In particular, the method only detects first-order leakage.
    We note, however, that this does not detract from \rosita achieving its
    aim of assisting in assessing compliance with the standard.
  \item \textbf{Model Limitations:} \rosita relies on \elmos which is only a model of
    the hardware.  Gaps between the model and the real hardware, both in terms of
    capturing the hardware behavior and in terms of accuracy of the model
    can result in missed leakage. 
    Moreover, hardware behavior may change over time, e.g.\ due to a microcode update~\cite{ReparazBV17}.
    For these reasons we recommend that operators do not rely solely on \rosita.
    It can be used to achieve a high degree of assurance, and is likely to automate
    a significant part of the work required for compliance,
    but testing with the real hardware is essential.
\end{itemize}

A further limitation of \rosita, which, like others, stem from the \elmos model is
its suitability for other processors.
\elmos uses a very simple model of the processor. It is suitable for small,
in-order, cacheless microcontrollers, such as the Cortex-M0, AVR processors such as
the ATMEGA328p, or small RISC-V processors.
However, the model is unlikely to be suitable for more advanced processors.
Nonetheless, we believe that the \rosita is important, because these microcontrollers
it targets are extremely popular in embedded devices, where they often
implement cryptographic functionalities.
At the same time, there is very little control of the physical environment
of such devices, allowing the attacker unfettered access and enabling the
type of attacks we defend against.
We leave porting \elmos and \rosita to other microcontrollers to future work.

Another direction to which \rosita could be extended is using other statistical
tests.  While the ISO~17825 does not require them, tests such as mutual information~\cite{gierlichs2008mutual}
and $\chi^2$~\cite{kelsey1998side}, have been proven useful as statistical distinguishers.
Moreover, extending \rosita to support second- and higher-order attacks would allow
supporting more secure implementations.
It may also be possible to extend \rosita to use correlation power analysis (CPA)~\cite{brier2004correlation}.
This would, however, require the operator to provide \rosita with possible values
to search for correlation.  It is not clear that this can be done in a generic fashion.

\section{Conclusions}

Since their introduction over two decades ago, physical side-channel attacks
have presented a serious security threat, particularly to small computational
devices that need to maintain secrets under the physical control of the adversary.
To protect against such attacks, many ciphers' implementations employ masking techniques that
combine intermediate values with randomly selected masks.
As a consequence, due to the mask being uniformly distributed,
leakage of a masked value does not reveal information to the adversary.
While proven secure against certain attacks, in practice masked implementations often leak
secret information due to unintended interactions between
masked values involving hardware they are loaded and stored to.
To fix these leaks, the common practice is to repeatedly ``tweak the code until it stops leaking''.

In this work, we have set out to explore if leakage emulators can be used for the \emph{automatic} elimination of side channel leakage from software implementations. 
To achieve this, we have created a code rewrite engine called \rosita and combined it with the extended leakage emulator \elmos:
\begin{itemize}
\item \rosita incorporates rules to mitigate leakage arising from operand interactions, register reuse, rotation operations, and memory operations.
\item \elmo has undergone a major upgrade to \elmos for two reasons: firstly,
  it had to be able to tell \rosita the cause of the leakage, and secondly, we
    have added support by including the values that instructions store in various
    micro-architectural storage elements, which hold state that can leak
    information.
\end{itemize}

In our proof-of-concept, we used \rosita with \elmos to automatically protect
masked implementations of AES, Xoodoo and ChaCha.  Our experiments using the actual
hardware show the absence of exploitable leakage at only a 64\% penalty to the
performance, which is the worst case i.e. ChaCha. As ChaCha is the most complex cipher of the three to mask (in terms of the overhead
in performance and resources) this is also reflected in the penalty in the tools' performance.

\ifAnon
\else
\section*{Acknowledgements}

Francesco Regazzoni received support from the European Union Horizon
2020 research and innovation program under CERBERO project (grant
agreement number {732105}).

This research was supported by 
the Australian Research Council 
project numbers DE200101577, DP200102364, and DP210102670;
the Research Center for Cyber Security at Tel-Aviv University established by the State of Israel, the Prime Minister's Office and Tel-Aviv University;
and by a gift from Intel.

\fi

  \bibliographystyle{IEEEtranSN}
{\footnotesize 

}
\newpage
\begin{appendices}
  \crefalias{section}{appendix}
\section{Validating the Setup}\label{app:elmovalid}
To validate our setup, we reproduce the results of \citet{McCann2017}.
Specifically, we perform a fixed vs.\ random test on the code in \cref{l:elmosnip},
which contains an implementation of one of the steps in the AES 
encryption known as the \textsc{ShiftRows} operation.
Specifically, register \texttt{r1} points to the 16 bytes that represent the state 
of the AES encryption.
\textsc{ShiftRows} performs a fixed permutation of these bytes.
The implementation loads three four-byte words and uses the \texttt{rors} instruction
to rotate the bytes, before storing them back to the state.

\begin{lstlisting}[numbers=none,caption=\textsc{ShiftRows} from~\citet{McCann2017},label=l:elmosnip,keepspaces]
ldr  r4, [ r1, #4 ]
rors r4, r5
str  r4, [ r1, #4 ]
ldr  r4, [ r1, #8 ]
rors r4, r6
str  r4, [ r1, #8 ]
ldr  r4, [ r1, #12 ]
rors r4, r3
str  r4, [ r1, #12 ]
\end{lstlisting}

For the fixed vs.\ random test we collect 2500 traces where the state contains fixed
data masked with the same mask value 
and 2500 traces where the state consists of random values masked with the same
mask.
A random value for the mask is chosen for each trace.
We compare the distribution of the power reading in each sample point between
the fixed and the random traces,
and calculate the Welch \ttest to check the likelihood that the two distributions
are the same.
As mentioned before, following common practice in side-channel analysis, we consider the distributions different enough to indicate
leakage if the absolute value of the \ttest value is above 4.5.

\begin{figure}[h]
	\centering
	\captionsetup[subfigure]{margin=10pt}
	\begin{subfigure}[t]{\columnwidth}
		\centering
		\includegraphics[width=\columnwidth]{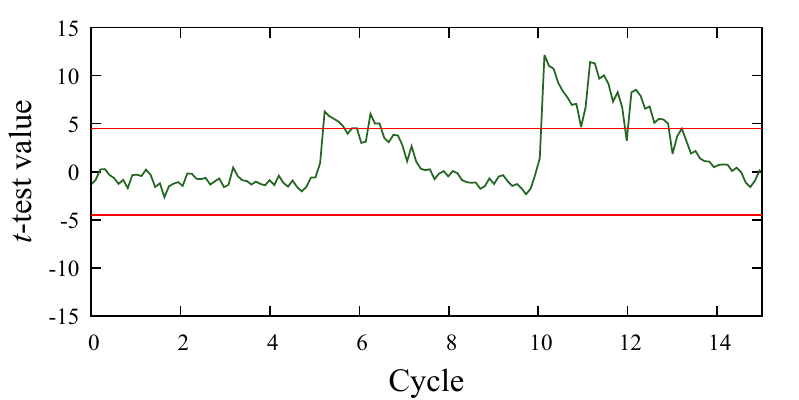}
		\caption{Real traces in STM32F030.}
		\label{f:sr_check}
	\end{subfigure}
	\begin{subfigure}[t]{\columnwidth}
		\centering
		\includegraphics[width=\columnwidth]{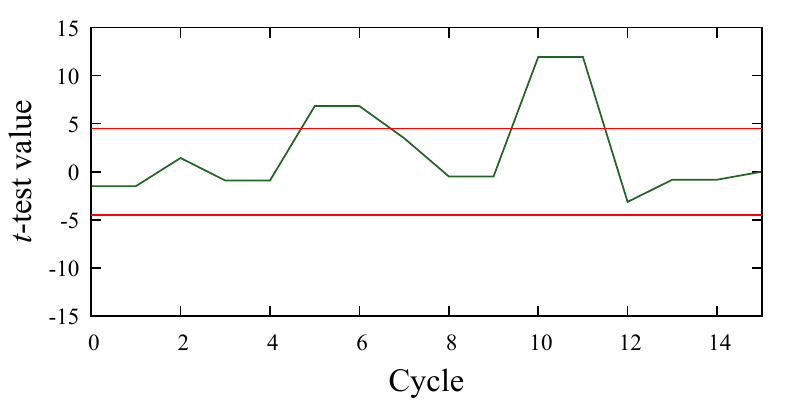}
		\caption{Simulated traces from \elmo.}
		\label{f:sr_check_elmo}
	\end{subfigure}
	\caption{Fixed vs.\ random of the AES \textsc{ShiftRows} operation.}
	\label{f:sr_compare_elmo}
\end{figure}

\cref{f:sr_check} shows the result of the fixed vs.\ random test.
The horizontal axis shows the time and the vertical axis shows the \ttest value.
We indicate instruction boundaries with vertical bars, and the \ttest threshold of $\pm4.5$ 
with horizontal red lines.
Comparing the figure to the results of running \elmo on the same code,
shown in \cref{f:sr_check_elmo}, we see that \elmo produces a fairly accurate
simulation of the leakage.

In particular, our figure resembles Figure 5 of \citet{McCann2017},
with only minor differences that reflect the different test environment.

\section{Eliminating Byte Interaction in Stores}\label{app:byteinteract}
\cref{f:storebytefix} shows an example of the rewrite rule for eliminating interactions in byte stores.
The code uses two registers chosen to not conflict with the store, 
\texttt{r0} and \texttt{r6} in the example in \cref{f:storebytefix}.
The first is used for selecting the byte to store, while the second is used for the byte.
\rosita uses two stores for each byte to avoid interactions on the memory bus or
in the DRAM.

\begin{figure}[htb]
  \begin{centering}
\begin{fix}
	  	
    str r2, [r3]
  \to
    push \{r6\}\linebreak
    push \{r0\}\linebreak
    movs r0, \#0xff\linebreak
    movs r6, r2\linebreak
    ands r7, r7\linebreak
    ands r6, r0\linebreak
    lsls r0, \#0\linebreak
    strb r0, [r3, \#0]\linebreak
    strb r6, [r3, \#0]\linebreak
    movs r6, r7\linebreak
    movs r6, r2\linebreak
    movs r0, \#0xff\linebreak
    lsls r0, \#8\linebreak
    ands r7, r7\linebreak
    ands r6, r0\linebreak
    lsrs r0, \#8\linebreak
    lsrs r6, \#8\linebreak
    strb r0, [r3, \#1]\linebreak
    strb r6, [r3, \#1]\linebreak
    .\linebreak
    .\linebreak
    .\linebreak
    pop \{r0\}\linebreak
    pop \{r6\}
\end{fix}
  \end{centering}
  \caption{Addressing byte interaction in stores. A leaking store instruction (left) and part of the fixed sequence (right).\label{f:storebytefix}}
\end{figure}

\section{Additional Figures}\label{app:figures}

\begin{figure*}[b]
  \begin{subfigure}{0.32\textwidth}
    \centering
    \includegraphics[width=0.9\textwidth]{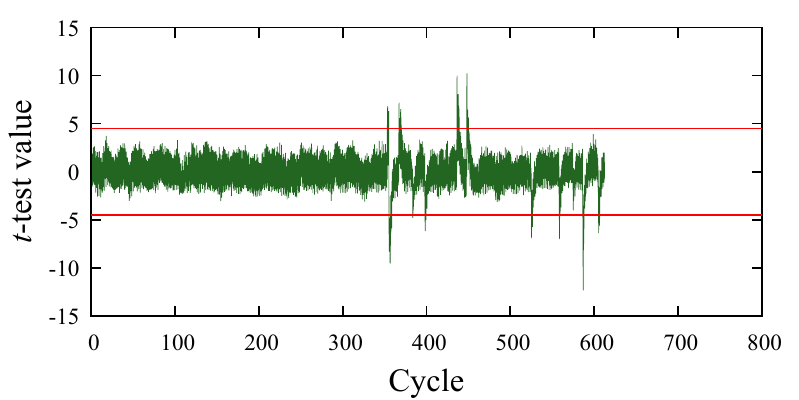}
    \caption{One fixed input -- before \rosita.\label{f:xoodoo-1b}}
  \end{subfigure}
  \begin{subfigure}{0.32\textwidth}
    \centering
    \includegraphics[width=0.9\textwidth]{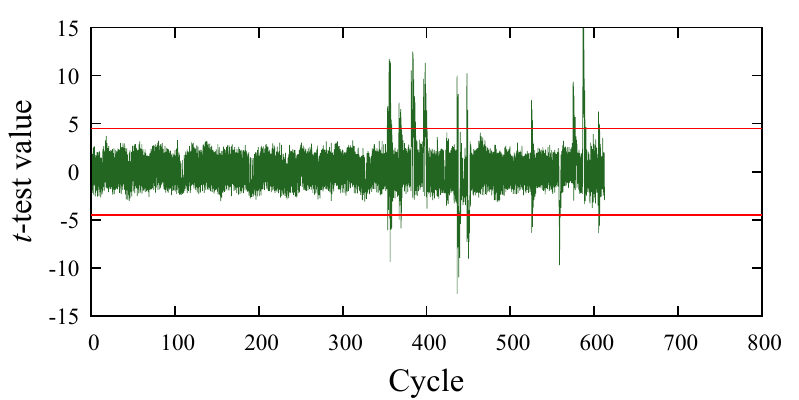}
    \caption{Five fixed inputs -- before \rosita.\label{f:xoodoo-5b}}
  \end{subfigure}
  \begin{subfigure}{0.32\textwidth}
    \centering
    \includegraphics[width=0.9\textwidth]{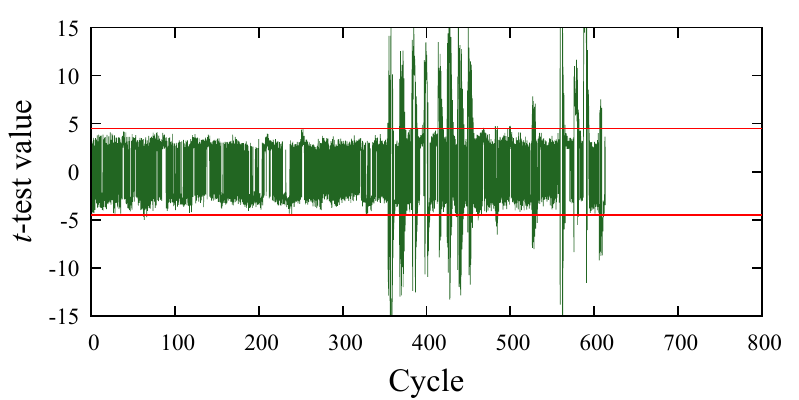}
    \caption{100 fixed inputs -- before \rosita.\label{f:xoodoo-100b}}
  \end{subfigure}\\
  \begin{subfigure}{0.32\textwidth}
    \centering
    \includegraphics[width=0.9\textwidth]{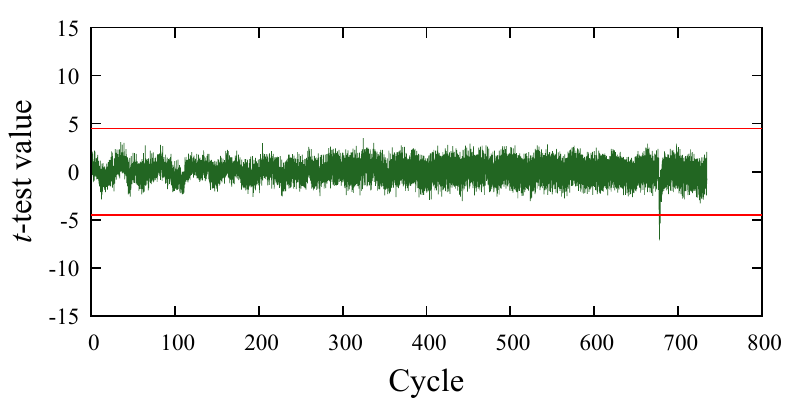}
    \caption{One fixed input -- after \rosita.\label{f:xoodoo-1a}}
  \end{subfigure}
  \begin{subfigure}{0.32\textwidth}
    \centering
    \includegraphics[width=0.9\textwidth]{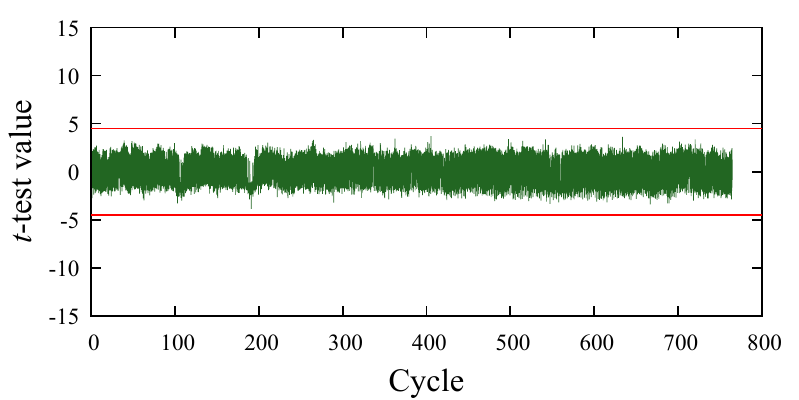}
    \caption{Five fixed inputs -- after \rosita.\label{f:xoodoo-5a}}
  \end{subfigure}
  \begin{subfigure}{0.32\textwidth}
    \centering
    \includegraphics[width=0.9\textwidth]{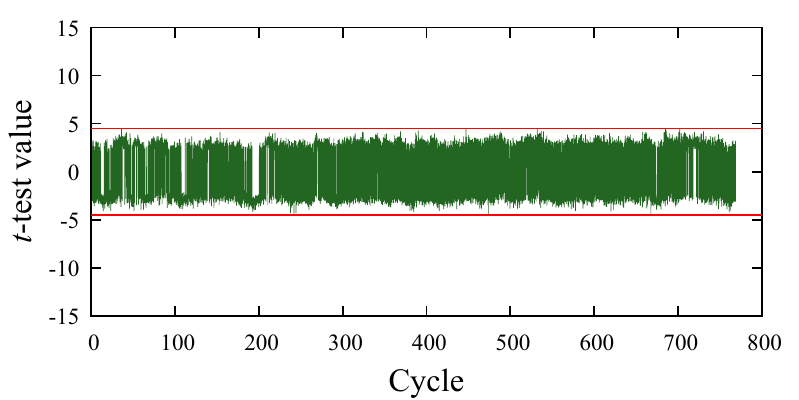}
    \caption{100 fixed inputs -- after \rosita.\label{f:xoodoo-100a}}
  \end{subfigure}
  \caption{$t$-test of masked Xoodoo implementation before and after \rosita, varying the number of fixed vs.\ random pairs.\label{f:Xoodoo-fix}}
\end{figure*}

\begin{figure*}[b]
	\begin{subfigure}{0.32\textwidth}
		\centering
		\includegraphics[width=0.9\textwidth]{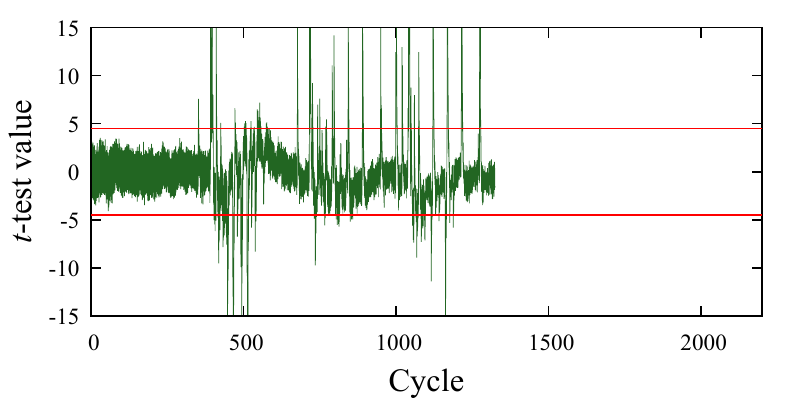}
		\caption{One fixed input -- before \rosita.\label{f:chacha20-1b}}
	\end{subfigure}
	\begin{subfigure}{0.32\textwidth}
		\centering
		\includegraphics[width=0.9\textwidth]{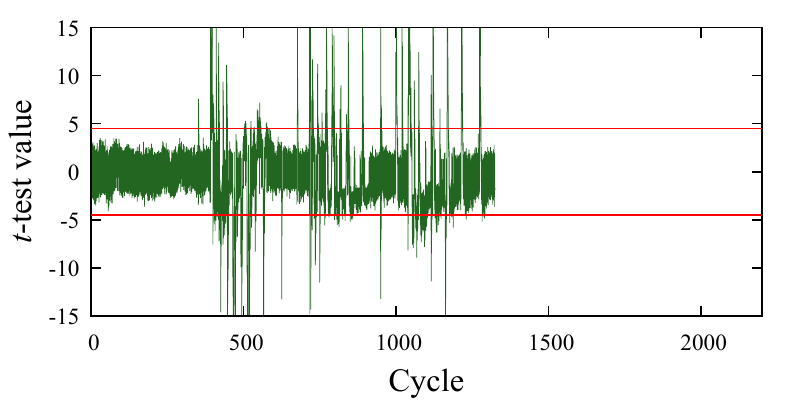}
		\caption{Five fixed inputs -- before \rosita.\label{f:chacha20-5b}}
	\end{subfigure}
	\begin{subfigure}{0.32\textwidth}
		\centering
		\includegraphics[width=0.9\textwidth]{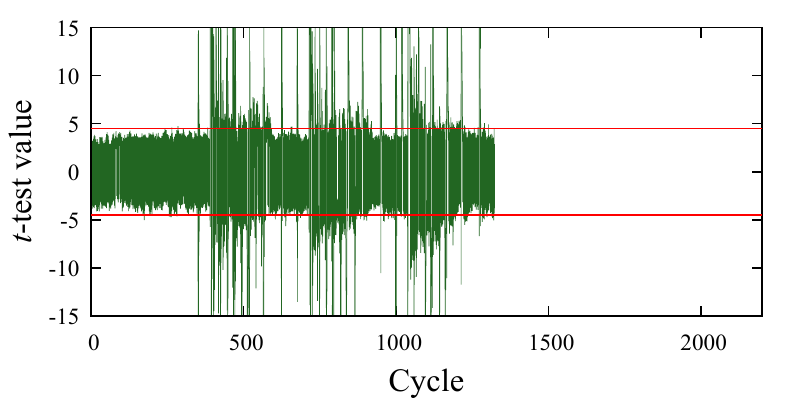}
		\caption{100 fixed inputs -- before \rosita.\label{f:chacha20-100b}}
	\end{subfigure}\\
	\begin{subfigure}{0.32\textwidth}
		\centering
		\includegraphics[width=0.9\textwidth]{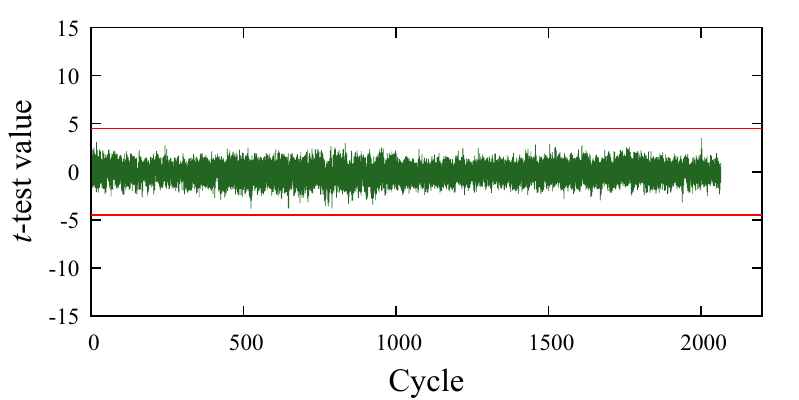}
		\caption{One fixed input -- after \rosita.\label{f:chacha20-1a}}
	\end{subfigure}
	\begin{subfigure}{0.32\textwidth}
		\centering
		\includegraphics[width=0.9\textwidth]{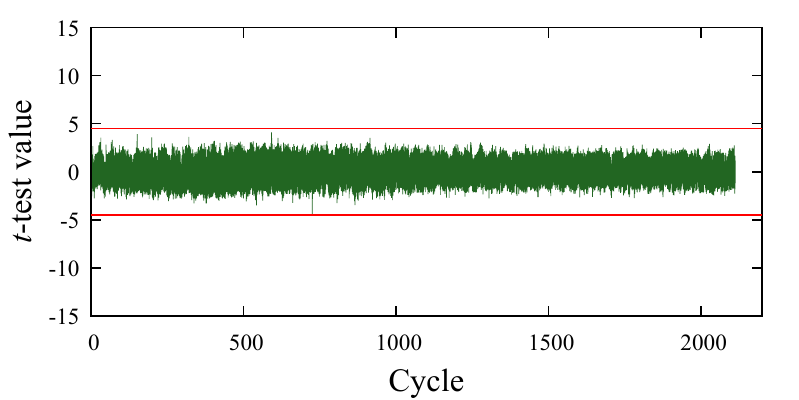}
		\caption{Five fixed inputs -- after \rosita.\label{f:chacha20-5a}}
	\end{subfigure}
	\begin{subfigure}{0.32\textwidth}
		\centering
		\includegraphics[width=0.9\textwidth]{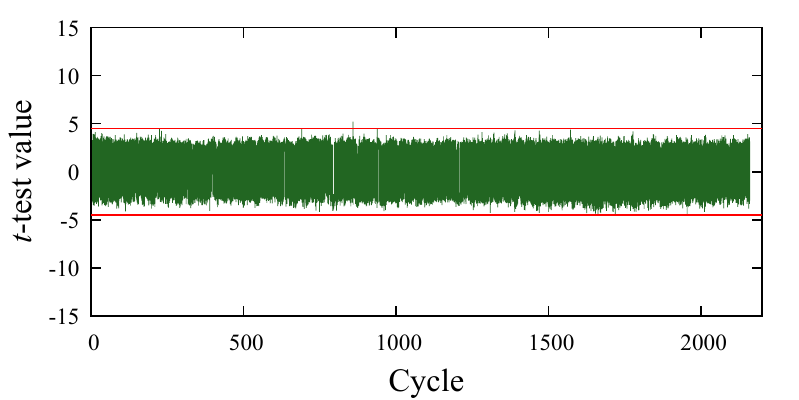}
		\caption{100 fixed inputs -- after \rosita.\label{f:chacha20-100a}}
	\end{subfigure}
	\caption{$t$-test of masked ChaCha implementation before and after \rosita, varying the number of fixed vs.\ random pairs.\label{f:chacha20-fix}}
\end{figure*}

\end{appendices}

\end{document}